\documentclass[11pt]{article}
\newcommand{\Comment}[1]{{}}
\usepackage{amsfonts,amsthm,amsmath,amssymb,slashed}
\usepackage[textwidth = 430 pt, textheight = 630 pt]{geometry}
\usepackage{color}

\Comment{\usepackage{color}
\definecolor{MyDarkBlue}{rgb}{0.15,0.15,0.45}
\usepackage[linktocpage=true]{hyperref}
\hypersetup{
colorlinks=true,
citecolor=MyDarkBlue,
linkcolor=MyDarkBlue,
urlcolor=MyDarkBlue,
pdfauthor={Horatiu Nastase and Jacob Sonnenschein},
pdftitle={The title},
pdfsubject={hep-th}
}

\usepackage[numbers,sort&compress]{natbib}
\usepackage{hypernat}}
\usepackage{graphicx}
\usepackage{cite}
\usepackage{url}

\newcommand\ignore[1]{}
\def\one{{\,\hbox{1\kern-.8mm l}}}

\def\Tr{{\rm Tr\, }}

\def\a{\alpha}\def\b{\beta}

\def\d{\partial}

\def\Tr{\mathop{\rm Tr}\nolimits}

\newcommand{\Cset}{{\,\,{{{^{_{\pmb{\mid}}}}\kern-.45em{\mathrm C}}}}}

\newcommand{\be}{\begin{equation}}
\newcommand{\bea}{\begin{eqnarray}}

\newcommand{\ee}{\end{equation}}
\newcommand{\eea}{\end{eqnarray}}

\newcommand{\non}{\nonumber \\}
\newcommand{\CR}{\non\cr}
\newcommand{\pa}{\partial}

\def\pa{\partial}

\newcommand{\aR}{{\alpha_R}}
\newcommand{\aI}{{\alpha_I}}
\newcommand{\bR}{{\beta_R}}
\newcommand{\bI}{{\beta_I}}

\begin{document}

\renewcommand{\thefootnote}{\fnsymbol{footnote}}

\makeatletter
\@addtoreset{equation}{section}
\makeatother
\renewcommand{\theequation}{\thesection.\arabic{equation}}

\rightline{}
\rightline{}



\begin{center}
{\LARGE \bf{\sc Novel knotted non-abelian gauge fields}} 
\end{center} 
 \vspace{1truecm}
\thispagestyle{empty} \centerline{
{\large \bf {\sc Horatiu Nastase${}^{a}$}}\footnote{E-mail address: \Comment{\href{mailto:horatiu.nastase@unesp.br}}
{\tt horatiu.nastase@unesp.br}}
{\bf{\sc and}}
{\large \bf {\sc Jacob Sonnenschein${}^{b}$}}\footnote{E-mail address: \Comment{\href{mailto:cobi@tauex.tau.ac.il}}{\tt cobi@tauex.tau.ac.il}}
                                                        }

\vspace{.5cm}


\centerline{{\it ${}^a$Instituto de F\'{i}sica Te\'{o}rica, UNESP-Universidade Estadual Paulista}} 
\centerline{{\it R. Dr. Bento T. Ferraz 271, Bl. II, Sao Paulo 01140-070, SP, Brazil}}
\vspace{.3cm}
\centerline{{\it ${}^b$School of Physics and Astronomy,}}
\centerline{{\it The Raymond and Beverly Sackler Faculty of Exact Sciences, }} 
\centerline{{\it Tel Aviv University, Ramat Aviv 69978, Israel}}
 
\vspace{1truecm}

\thispagestyle{empty}

\centerline{\sc Abstract}

\vspace{.4truecm}

\begin{center}
\begin{minipage}[c]{380pt}
{\noindent

In analogy to null electromagnetic fields we define null YM fields.
We show that the null non-abelian $SU(N)$ gauge  fields admit a set of $2 N^2$ conserved ``helicities''.
We derive  null YM  solutions that carry finite helicities by uplifting the abelian Hopfion solution and their generalizations. 
Another method that we implement is to deform YM solutions which do not carry helicities into ones that have 
nontrivial helicities. A nontrivial non-Abelian solution with helicities is found as a wave of infinite energy. 
We also discuss  non-abelian  generalizations of the Bateman parameterization for null abelian gauge fields.

}
\end{minipage}
\end{center}

\vspace{.5cm}

\setcounter{page}{0}
\setcounter{tocdepth}{2}

\newpage

\tableofcontents
\renewcommand{\thefootnote}{\arabic{footnote}}
\setcounter{footnote}{0}

\linespread{1.1}
\parskip 4pt




\section{Introduction}

Despite being a linear theory, Maxwell electromagnetism in vacuum 
admits ``knotted" solutions, with a conserved topological charge (the Hopf index). The basic knot is  the electromagnetic Hopfion solution found in 
\cite{Ranada:1989wc,ranada1990knotted}, following earlier work of \cite{Trautman:1977im}. Generalization  of the Hopfion were writtten down   in \cite{Arrayas:2015aa,RanadaBook} and 
\cite{besieris2009hopf,Kedia:2013bw,Hoyos:2015bxa} (see the review \cite{arrayas2016knots}
for more details and a more complete list of references). 
These solutions are space and time dependent, and have also nonzero ``helicities" ${\cal H}_{rs}=\int d^3x \epsilon^{ijk}A_i^{(r)}
\d_j A_k^{(s)}$, with $r,s=e,m$ (electric
and magnetic), written as Chern-Simons forms of $A^{(r)}_i$ = spatial gauge field $A_i$ and Poincar\'{e} dual gauge field $C_i$. 

The simplest way to formulate the knotted solutions  is in terms of the Bateman formulation \cite{Bateman:1915}, 
where the Riemann-Silberstein vector $\vec{F}=\vec{E}+i\vec{B}$ is written in terms of complex scalars $\a,\b\in\mathbb{C}$
as $\vec{F}=\vec{\nabla}\a\times \vec{\nabla}\b$, in which case the Hopfions are null fields, $\vec{F}^2=0$. The 
null condition ensures the conservation in time of the helicities ${\cal H}_{r,s}$.
In \cite{besieris2009hopf,Kedia:2013bw} it was shown that one can find new solutions by replacing the pair $(\a,\b)$ with 
holomorphic functions of them, $\tilde \a=f(\a,\b),\tilde \b=g(\a,\b)$, and in particular  one can 
obtain ``$(p,q)$-knotted solutions" by using the transformation $\a\rightarrow \a^p,\b\rightarrow \b^q$ on the basic 
topological solution, 
with ${\cal H}_{ee}={\cal H}_{mm}=1$.

A simple method to derive the topological  non-trivial solutions from  topologically trivial ones was found in 
\cite{Hoyos:2015bxa}. It was proven there that  
acting with a special conformal transformation (SCT) with {\em complex} rather than real parameter  on a null solution 
yields a different null solution (it is a solution-generating technique, rather than a symmetry). 
Moreover, it was  shown that applying a SCT with the parameter of transformation 
$b=(i,0,0,0)$ on the trivial solution of constant electric and magnetic fields that are perpendicular and have the same 
absolute value, yields the Hopfion solution. Note that this SCT maps a topologically trivial solution with infinite energy 
to a topologically non-trivial solution with finite energy.  Similar knotted solutions were derived also by applying such
 transformation on plane wave solutions. 

The non-trivial topological EM solutions were mapped in \cite{Alves:2017zjt} into knotted solution of hydrodynamics. In 
\cite{Nastase:2022aps} it was  proven that the Hopfion fluid-electromagnetic knot, carrying fluid and electromagnetic
helicities, solves the fluid dynamical equations as well as the Abanov Wiegmann  equations\cite{Abanov:2021hio} 
\footnote{For earlier work considering chiral liquids, i.e., liquids with chirality with respect to electromagnetism, and the conservation of helicities in this context, see \cite{Avdoshkin:2014gpa,Kirilin:2017tdh}.}
for the conservations of the fluid  helicities. 

 A natural question, which we attempt to answer in this paper, is how much of this can be extended to Yang-Mills theories?
We will use a natural generalization of the Riemann-Silberstein vector $\vec{F}$, to define both singlet and $T^a$-dependent 
null fields, and corresponding helicities ${\cal H}^{NA}_{r,s}, {{\cal H}^{NA}}^a_{r,s}$, in total 2$N^2$ 
of them for the group $SU(N)$, the $(e,e)$ and $(m,m)$ ones being
conserved in time for the corresponding null fields. We will then construct solutions with nonzero such helicities, in three ways:
either by embedding the abelian ones into the non-abelian ones, by using {\em a priori} non-abelian ansatz, or by 
deforming solutions with zero helicities. Finally, we will give two possible  generalizations of the Bateman 
formulation for a non-abelian group ${\cal G}$, 
with $(\a,\b)$ replaced by scalar fields $({\cal A}, {\cal B}\in {\cal G})$. 

We  summarize now the properties of the   solutions we found  that admit
topological non-trivial helicities.
\begin{itemize}
\item
Needless to say that  they are exact solutions   
 of Maxwell or Yang-Mills equations.
\item 
The Abelian Hopfion has gauge field that is smooth and nonsingular, 
with finite energy, 
and has a nonzero helicity which equals the  Hopf index, 
but by a gauge transformation large enough at infinity, it can 
become of zero helicity.
\item
The basic (1,1)  Hopfion solution can be generalized to a set of infinitely many solutions characterized by (m,n) $m,n\in {\cal Z}$. 
\item
The Abelian Hopfion and its generalizations  can be uplifted to  solutions with nonzero non-Abelian singlet helicities.
\item
A radial non-Abelian time-dependent
wave can have nonzero and finite non-Abelian helicity 
 but it is singular and has an infinite energy.
\item
A non-Abelian time-dependent plane wave has nonzero and finite non-Abelian helicity, but it is singular and 
has infinite energy.
\item
The Abelian monopole can be gauge transformed to a solution that has finite helicity, though not interpretable 
as Hopf index, which has finite energy and is static, but is singular at the origin.
\item
The non-Abelian monopole can be gauge transformed to a solution with finite non-Abelian helicity, interpretable 
as Hopf index, static and with finite energy, and nonsingular. 
\item
Note that non of these fields are Beltrami fields, since $\vec{B}=\vec{\nabla}\times \vec{A}$ is not parallel 
to $\vec{A}$ ($\vec{A}$ has a piece satisfying $\vec{A}\cdot \vec{B}=0$ in all cases).
\item
 We address here  the classical solutions. The issue of  the quantization 
of the helicity in quantum theory is tricky, and do not attempt to address  it here.
\item
 The quantized helicity in fact would be 
a Hopf index, and we have not been able to define the non-Abelian case for the Hopf index.
\item
We will attempt a non-Abelian Bateman formulation for the field strengths, which is not like the Clebsch 
or Darboux parametrization: it is not for the gauge fields themselves, and it assumes part of the equations 
of motion.
\end{itemize}

The paper is organized as follows. In section 2 we review the electromagnetism (abelian) case, for null fields, helicities and 
Hopfion. In section 3 we extend the notion of null fields and of helicities to Yang-Mills fields. In section 4 we consider solutions
with nonzero helicities, first uplifting the abelian solutions, then considering a radial non-abelian wave ansatz (with infinite 
energy), then deforming trivial solutions to ones that admit finite helicities, and finally describing two proposals for the 
non-abelian generalization of the Bateman formulation. In section 5 we conclude and present open questions.

\section{Abelian null fields and helicities}

\noindent

We begin by reviewing  the notion of   "null" electromagnetic fields, expressed in terms of the  complex Riemann-Silberstein vector field 
$\vec{F}$,  and the  Bateman parameterization of  the null fields  in terms of two complex scalar fields $\alpha$ and $\beta$. 
After writing Maxwell's equations  in terms of these fields, we write down
several simple null solutions in terms of the Bateman parametrized fields.  

In the next subsection,  we review the fact that  the null  fields admit four conserved charges,  referred to  as  helicites.  
We denote them as $({\cal H}_{mm}, {\cal H}_{me}, {\cal H}_{em},{\cal H}_{ee})$.
We then calculate these conserved charges for the examples of the null fields.

\subsection{Null Electromagnetic fields}\label{nullEMsol}

\noindent

The electromagnetic fields $\vec E$ and $\vec B$ can be combined into a    complex vector, the so-called
Riemann-Silberstein vector,
\be
\vec{F}=\vec{E}+i\vec{B}\;.
\ee

The vacuum Maxwell's equations of motion in terms of $\vec{F}$ are 
\be
\vec{\nabla}\cdot\vec{F}=0\;;\;
\d_t\vec{F}+i\vec{\nabla}\times \vec{F}=0. \label{MxF}
\ee

One can then define null fields by
\be\label{nullEB}
F^2=0 \qquad \leftrightarrow \qquad\vec E\cdot \vec B=0 \qquad E^2-B^2=0.
\ee

A useful parametrization of the null fields  is in terms of  the Bateman ansatz \cite{Bateman:1915},
\be
\vec{F}=\vec{\nabla}\a\times\vec{\nabla}\b\;,
\ee
in terms of the two complex scalar fields $\a,\b\in\mathbb{C}$. 

In components, the ansatz is 
\bea
E^i&=& \epsilon^{ijk} \left(\pa_j\aR\pa_k\bR- \pa_j\aI\pa_k\bI\right)\cr
B^i&=& \epsilon^{ikj} \left(\pa_j\aR\pa_k\bI+ \pa_j\aI\pa_k\bR\right)\;,\label{compEB}
\eea
where the indices $I$ and $R$ refer to the imaginary and real parts, respectively.

The first  Maxwell equation  in (\ref{MxF})  is  trivially satisfied by the Bateman ansatz and the second takes the form 
\be
i\nabla\times (\pa_t\alpha\nabla\beta-\pa_t\beta\nabla\alpha)=\nabla\times \vec F\;,\label{FMx}
\ee
which is satisfied if
\be
i(\pa_t\alpha\nabla\beta-\pa_t\beta\nabla\alpha)=\vec F.
\ee

In fact in a more general solution the right-hand side could be 
$\vec F +\vec G$, where $\nabla\times \vec G=0$, but $\vec G$ can be time dependent.

We note that {\em on-shell}, we can write 
\be
A_\mu= \frac12{\rm Im}[ \alpha\pa_\mu \beta- \beta\pa_\mu \alpha]\;.
\ee

Solutions of (\ref{FMx}) have necessarily a zero norm, since 
\be\label{null}
F^2=(\pa_t\alpha\nabla\beta-\pa_t\beta\nabla\alpha)(\vec{\nabla}\a\times\vec{\nabla}\b)=0 \;.
\ee

The simplest null electromagnetic field is made up of perpendicular constant electric and magnetic fields, namely
\be\label{basicEM}
\vec E = (E,0,0)\;, \qquad \vec B = (0,B,0)\;,\qquad |E|=|B|.
\ee

In terms of $\a$ and $\b$, for $E=-B=-4$, this solution takes the form
\be\label{Bateconstant}
\a= 2i(t+z)-1\;, \qquad \b=2(x-iy).
\ee

Another class of well known null electromagnetic solutions are the plane waves
\be
\vec F = (\hat x +i \hat y) e^{i(z-t)}\;,
\ee
which in terms of the Bateman variables read
\be
\alpha= e^{i(z-t)}\;,\qquad \beta = x+ i y.
\ee

The basic knotted solution, the Hopfion solution 
is given by
\be
\a=\frac{A-iz}{A+it}\;,\;\;
\b=\frac{x-iy}{A+it}\;,\;\;
A=\frac{1}{2}(x^2+y^2+z^2-t^2+1).\label{absol}
\ee

The Hopfion is 
characterized by non-trivial helicity  where ${\cal H}_{mm} =\frac14$ where  ${\cal H}_{mm}$, is  defined below in (\ref{Hmm}).    

This solution can be obtained by applying the  special conformal transformation (SCT)
\be
x^\mu \rightarrow \frac{x^\mu -b^\mu x_\sigma x^\sigma}{1- 2 b_\sigma x^\sigma + b_\rho b^\rho x_\sigma x^\sigma}
\ee
on (\ref{Bateconstant}), with $b^\mu =(i,0,0,0)$.

\subsection{Conserved ``Helicities''}

We now introduce a set of four quantities refered to as helicities,  that are conserved in time for null electromagnetic fields. 
 We use the Coulomb gauge 
\be
A_0=0 \;,
\ee
so that
\be
E^i=F^{i0}=-\pa^0A^i\;, \qquad F^{ij}=\epsilon^{ijk}B_k\;,\qquad \vec B=\vec \nabla\times\vec A .
\ee

The first helicity, referred to as (magnetic, magnetic) is given by
the Chern Simons form  
\bea\label{Hmm}
{\cal H}_{mm}&=&\int d^3 x d A \wedge A= \int d^3x\epsilon^{ijk}A_i\d_j A_k\CR
&=&\int d^3x \vec{A}\cdot \vec{B} .
\eea 

This helicity is conserved in time, provided that $\vec{E}\cdot\vec{B}=0$, since from the Maxwell's equation $\d_t \vec{B}
=-\vec{\nabla}\times \vec{E}$ and a partial integration, we obtain
\be
\d_t {\cal H}_{mm}=\int d^3x (\d_t \vec{A}\cdot \vec{B}+\vec{A}\cdot \d_t\vec{B})
=-\int d^3x(\vec{E}\cdot\vec{B}+\vec{A}\cdot(\vec{\nabla}\times\vec{E}))= -2\int d^3x \vec{E}\cdot\vec{B}.
\ee

In a similar manner, also  the helicity ${\cal H}_{me}$, defined by 
\be
{\cal H}_{me}=\int d^3x \vec{A}\cdot\vec{E}\;,
\ee
is conserved, provided that $(\vec{E}^2-\vec{B}^2)=0$,  since by using the Maxwell's equation $\d_t\vec{E}=\vec{\nabla}\times
\vec{B}$ and a partial integration, we get 
\be
\d_t {\cal H}_{me}=\int d^3x(\d_t \vec{A}\cdot \vec{E}+\vec{A}\cdot\d_t\vec{E})
=\int d^3x(-\vec{E}\cdot\vec{E}+\vec{A}\cdot(\vec{\nabla}\times\vec{B}))
=-\int d^3x(\vec{E}^2-\vec{B}^2)\;.
\ee

Therefore for null fields both ${\cal H}_{mm}$ and ${\cal H}_{me}$ are conserved.

For Maxwell fields in the vacuum (i.e., without sources), one can also consider the formulation in terms of the 
Poincar\'{e} dual field $C$, so
\be
\tilde F_{\mu\nu}=\frac{1}{2} \epsilon_{\mu\nu\rho\sigma} F^{\rho\sigma}\equiv (d C)_{\mu\nu}
=\pa_\mu C_\nu- \pa_\nu C_\mu \qquad \rightarrow \qquad \vec E =\vec \nabla\times \vec C\;,
\ee
in the gauge $C_0=0$, dual to the Coulomb gauge.
Using together the basic and dual formulation (which is possible in the absence of sources), 
the helicity ${\cal H}_{me}$   can also be written as a (mixed) Chern-Simons term (or rather, BF term),
\be
{\cal H}_{me}=\int d^3x \vec{A}\cdot\vec{E}=\int d^3x \epsilon^{ijk}A_i\d_j C_k.
\ee

In addition to ${\cal H}_{mm}, {\cal H}_{me}$, we can now define two additional helicities, ${\cal H}_{ee} {\cal H}_{em}$,  
given by the dual field CS form
\be
{\cal H}_{ee}=\int d^3x \vec{C}\cdot \vec{E}=\int d^3x \vec{C}\cdot \vec{\nabla}\times \vec{C}
=\int d^3x \epsilon^{ijk}C_i\d_j C_k\;,\label{Hee}
\ee
and the BF form
\be
{\cal H}_{em}=\int d^3x \vec{C}\cdot\vec{B}=\int d^3x \epsilon^{ijk}C_i\d_j A_k\;,
\ee
which are  conserved for null configurations, as follows from 
\be
\d_t {\cal H}_{ee}=\int d^3x (\d_t \vec{C}\cdot \vec{E}+\vec{C}\cdot \d_t\vec{E})
=-\int d^3x(\vec{B}\cdot \vec{E}+\vec{C}\cdot (\vec{\nabla}\times \vec{B}))\;,
\ee
\be
\d_t {\cal H}_{em}= \int d^3x(\d_t \vec{C}\cdot \vec{B}+\vec{C}\cdot \d_t \vec{B})
=\int d^3x(-\vec{B}\cdot \vec{B}+\vec{C}\cdot (\vec{\nabla}\times \vec{E}))
=\int d^3x (\vec{E^2}-\vec{B}^2)\;\;,
\ee
again by use of the Maxwell's equations and partial integrations. 

Then, note that an essential point about the conservation of the helicities is that the corresponding fields must:
1) be on-shell (solutions of the Maxwell's equations); and 2) go to zero sufficiently fast at infinity such that the boundary 
terms that we dropped when partially integrating vanish. For instance, for time invariance of ${\cal H}_{mm}$, we need that 
\be
\oint_{\Sigma_\infty}dS_j\epsilon^{ijk}A_i E_k=0. 
\ee

But, since the helicities are written in terms of the gauge fields, we need also to consider their possible gauge dependence, 
which is important, as we will see later. As CS forms, they are only gauge invariant provided relevant boundary terms 
vanish also. For instance, for the ${\cal H}_{mm}$ helicity, we have 
\be
\delta_{\rm gauge}{\cal H}_{mm}=\delta_{\rm gauge} 
\int_V  A\wedge d A=\int_V  F\wedge d\lambda=\int_V d(F\lambda)=\oint_{\rm bd.} F\lambda\;,
\ee
so it is gauge invariant only if the above product of the field strength $F_{\mu\nu}$ and the gauge parameter $\lambda$ 
integrates to zero on the boundary at infinity. Otherwise, ${\cal H}_{mm}$ could be gauge dependent. 

To summarize, null electromagnetic configurations, $\vec F^2=0$, admit the conservation in time of the four helicities  
\be
{\cal H}_{mm},\qquad {\cal H}_{me},\qquad {\cal H}_{em},\qquad {\cal H}_{ee} .
\ee

In fact, for the conservation of ${\cal H}_{mm}$ and ${\cal H}_{ee}$ it is suffices to have $\vec E\cdot \vec B =0$, 
and for the conservation of ${\cal H}_{me}$ and ${\cal H}_{em}$, one only needs $\vec E^2-\vec B^2=0$.

For the basic solution of perpendicular,  constant, and equal electric and magnetic fields, given by (\ref{basicEM}), 
all the helicities vanish.

On the other hand, for the Hopfion solution of (\ref{absol}), the conserved helicities are 
\be
{\cal H}_{mm}=\frac14,\qquad {\cal H}_{me}=0,\qquad {\cal H}_{em}=0,\qquad {\cal H}_{ee}=\frac14 .
\ee

Several other classes of topologically non trivial null EM fields were derived in \cite{Hoyos:2015bxa}.

\subsection{${\cal H}_{mm}$ helicity and the Hopf index}

\noindent

The Hopfion solution has not only finite ${\cal H}_{mm}$, but also {\em finite energy}
and a finite Hopf index, a topological number. In fact, it is related to ${\cal H}_{mm}$ for the case of the electromagnetic 
Hopfion. 

The first observation is that ${\cal H}_{mm}$, being a CS form in the 3 spatial dimensions, is only gauge invariant 
for a product of field strengths and gauge transformation parameters that integrates to zero at infinity, as we saw above, 
so it can only be identified with a topological number in that case. Secondly, time invariance also imposed a condition 
on the field strengths times gauge fields at infinity, also as we saw above. 

The Hopf index is defined for more general solutions than electromagnetic ones. 

In fact, rather generally, a Hopfion solution is defined by the complex scalar function in 3 (spatial) dimensions
\be
\phi_{\rm Hopfion}(x,y,z)=\frac{2(x+iz)}{2z+i(r^2-1)}.
\ee

This solution goes to 0 at infinity {\em in all directions} (so there is effectively a {\em single} point at infinity), 
so it effectively compactifies the space, since now we can add the point at infinity to the open $\mathbb{R}^3$ space
and obtain $S^3=\mathbb{R}^3(x,y,z)\cup \{\infty\}$ (for instance through the stereographic projection in 4 Euclidean dimensions). 

Furthermore, we can add a gauge field defined in terms of $\phi_{\rm Hopfion}$ and define an electromagnetic Hopfion 
solution. Indeed, define (for $\phi=\phi_{\rm Hopfion}$)
\be
{\cal F}=\frac{1}{2}F_{ij}dx^i\wedge dx^j=\frac{1}{4\pi i}\frac{\d_i\bar \phi\d_j\phi-\d_j\bar\phi\d_i\phi}{(1+|\phi|^2)^2}dx^i\wedge
dx^j\;,
\ee
then the Hopfion number, or index, can be identified with ${\cal H}_{mm}$ (but on the $S^3$, which 
comes from $\mathbb{R}^3$ with the
point at infinity!), since 
\be
N_{\rm Hopf}=\int_{S^3}{\cal A}\wedge {\cal F}={\cal H}_{mm}(S^3)
\ee
is now a topological number, meaning it is independent of small variations in the fields or the space.

Note that one solution to ${\cal F}=d{\cal A}$ is 
\be
A_i^{(0)} =\frac{1}{4\pi i}\frac{\bar\phi \d_i \phi-\phi\d_i\bar\phi}{1+|\phi|^2}\;,
\ee
as one can check. 

Further, replacing $\phi=\phi_{\rm Hopfion}$, one finds 
\be
A_i^{(0)}=\frac{2}{\pi } \frac{[2xz+z(r^2-1)]\d_i [2z^2-x(r^2-1)]-[2z^2-x(r^2-1)]\d_i [2xz+z(r^2-1)]}{[4z^2+(r^2-1)^2][4r^2+(r^2-1)^2]}.
\ee

We note that at infinity, generically we have $\phi\sim 1/r$, $A_i\sim 1/r^3$, so $E_i,B_i\sim 1/r^4$, 
whereas at $r\sim 0$, we have $\phi\sim 0$, 
$A_i \sim 0$, so $E_i,B_i\sim 0$, so the field is smooth at the origin and vanishes sufficiently fast at infinity 
to guarantee that: 1) there is a single point at infinity, so space is compactified, and 2) the boundary terms in the 
variation of the helicities vanish, so the helicities ${\cal H}$ are time independent.

Further, the magnetic field is then
\be
B_i =\frac{\epsilon_{ijk}}{2\pi i}\frac{\d_j \phi \d_k\bar\phi}{(1+\bar\phi\phi)^2}.
\ee

But we can also rewrite the gauge field as 
\be
A_i^{(0)} =\frac{1}{4\pi i}\left[\d_i \ln (1+\bar\phi\phi)-\frac{2\phi\d_i\bar\phi}{1+\bar\phi\phi}\right]
=\frac{1}{4\pi i}\left[\d_i \ln (1+\bar\phi\phi)\right]+A_i^{(1)}
\;,
\ee
so the second term, $A_i^{(1)}$, gives explicitly zero in $\vec{A}\cdot \vec{B}$, whereas the first (a pure gauge term!) 
can give a nonzero contribution
only due to the "singularity" of this gauge transformation at $r=\infty$ (it is a large enough gauge transformation at infinity). 
Indeed, as we saw, we obtain that $N_{\rm Hopf}={\cal H}_{mm}(\phi_{\rm Hopfion})\neq 0$.

\section{ Non-abelian null fields and helicities}

\noindent 

Following the derivation of the helicities and their conservations for null abelian fields,
 we introduce the analogous non-abelian constructions.

\subsection{Null Yang-Mills fields}

\noindent

Consider the Yang-Mills action in Minkowski space ($\Tr[T^a T^b]=\delta^{ab}$),
\be
S_{\rm YM}= -\frac14\int d^4 \Tr [F_{\mu\nu}F^{\mu\nu}]= -\frac14 \int d^4 F^a_{\mu\nu}F_a^{\mu\nu}\;,
\ee
where $F_{\mu\nu}= F^a_{\mu\nu}T^a$,  and $T^a$ are matrices in the adjoint representation for some group, say $SU(N)$, 
such that 
\be
[T^a,T^b]= f^{abc} T_c
\ee
and 
\be
F^a_{\mu\nu}= \pa_\mu A^a_\nu - \pa_\nu A^a_\mu + g f^{abc} A^b_\mu A^c_\nu.
\ee

The corresponding equations of motion are 
\be
D^\mu F_{\mu\nu}= 0:\qquad \pa^\mu F^a_{\mu\nu} + g {f^a}_{bc}A^{b\mu} F_{\mu\nu}^c=0.
\ee

Here also we use the Coulomb-like gauge
\be
A^a_0=0\;,
\ee
and the chromo-electric and magnetic fields are given by
\be
E^a_i=F^a_{i0}=-\pa_0 A^a_i\;, \qquad F^a_{ij}=\epsilon_{ijk} (B^a)^k.
\ee

We define the non-abelian complex vector
\be
\vec F= \vec F^a T^a = (\vec E^a+ i \vec B^a)T^a.
\ee

One can now define several null vectors. The first one, which is an obvious generalization of the abelian case, is 
\bea
&&\Tr[(\vec F)^2] =0 \qquad \rightarrow \qquad \frac12\left( \vec E^a \cdot \vec E^a- \vec B^a \cdot\vec B^a) 
+i (2 \vec B^a\cdot \vec E^a\right )=0\CR
&&\qquad \rightarrow \qquad ( \vec E^a \cdot \vec E^a- \vec B^a \cdot\vec B^a)=0 \;\;\;{\rm and}\;\;\; \vec B^a\cdot \vec E^a=0\;,
\eea
 which will be referred as the ``unit null'', or singlet null.
It is straightforward to generalize the null condition to  the following 
\bea 
&&\Tr [T^a (\vec F)^2]= [( \vec E^b \cdot \vec E^c- \vec B^b \cdot\vec B^c) +i  ( \vec E^b \cdot \vec B^c+\vec E^b \cdot
\vec{B}^c)]\Tr[T^aT^bT^c]=0 \qquad \rightarrow \cr
&&\qquad
d^{abc}( \vec E^b \cdot \vec E^c- \vec B^b \cdot\vec B^c)=0 \qquad
 d^{abc}( \vec E^b \cdot \vec B^c)=0\;,
\eea
which will refer to as the ``$T^a$ null'' fields.


\subsection{Helicities of non-abelian gauge fields}

\noindent

In analogy to the abelian gauge theory, we propose  that non-Abelian CS terms integrated over space are defined as 
helicities, and check under what conditions they are conserved in time. 

In the Abelian case, the helicities were interpreted also as linking number for flux lines, so the "knottedness" of solutions 
was classified by them. In the non-Abelian case, it is not clear if a similar concept can be introduced.

\subsubsection{Singlet $mm$ non-abelian helicity}

\noindent

The first (singlet) helicity is the analog of the ${\cal H}_{mm}$ one, and 
takes the form
\bea
{\cal H}^{NA}_{mm}&=& \int d^3 x \Tr\left[ A\wedge dA+ \frac23 A\wedge A \wedge A\right]=\int d^3x \Tr\left[A\wedge F 
-\frac{1}{3}A\wedge A\wedge A\right]\cr
&=&\int d^3 x \left [ A^a_i B^a_i 
-\frac13 \epsilon^{ijk}f^{abc}A^a_iA^b_jA^c_k\right ].\label{helNA}
\eea

The conservation of the helicity requires the vanishing of 
\bea
\d_t {\cal H}^{NA}_{mm} &=&\int d^3 x \left [ \d_tA^a_i B^a_i + A^a_i \d_tB^a_i - \epsilon_{ijk}f^{abc}
\d_tA^a_iA^b_jA^c_k\right ]\CR
&=& \int d^3 x \left [ E^a_i B^a_i + A^a_i(\d_t B^a_i -
\epsilon_{ijk}f^{abc}\d_tA^b_jA^c_k) \right ].
\eea

Next we make use of 
\bea
B^a_i&=& \frac12 \epsilon_{ijk} F^{ajk}= \frac12 {\epsilon_{i}}^{jk}(\pa_j A^a_k
-\pa_k A^a_j + {f^a}_{bc} A^b_j A^c_k)\cr
E^a_i&=& F^a_{i0}=\d_i A_0^a-\d_t A_i^a+{f^a}_{bc}A_i^b A_0^c=-\d_t A_i^a\;,\label{EBa}
\eea
where the last form is in the $A_0^a=0$ gauge, 
and the YM equation
\be
\d_t B^a_i + (\vec D\times \vec E)^a_i=\d_t B^a_i + \epsilon_{ijk}(\d_j E^a_k + f^{abc} A^b_j E^c_k)=0\;,\label{YMB}
\ee
and  after integration  by parts  we get that
\be
\d_t {\cal H}^{NA}_{mm} =2 \int d^3 x \vec E^a\cdot \vec B^a.
\ee

This implies that for the singlet null non-abelian configuration, for which 
\be
\vec E^a\cdot \vec B^a=0\;,
\ee
the helicity ${\cal H}^{NA}_{mm}$ is conserved.


\subsubsection{Dual and mixed singlet helicities}

\noindent

In the abelian gauge fields theory we have identified the helicity 
${\cal H}_{ee}$ by using the dual gauge field $C_\mu$. In the gauge $C_0=0$ the helicity took the form (\ref{Hee}). 
In the case of non-abelian gauge fields one can introduce a non-abelian dual gauge field
$\vec C^a$, in the gauge $C_0^a=0$, 
and in analogy to (\ref{Hee}) write down the dual singlet helicity (chromo-electric-chromo-electric)
${\cal H}^{NA}_{ee}$,
\bea
{\cal H}^{NA}_{ee}&=& \int d^3 x \Tr\left[ C\wedge dC+ \frac23 C\wedge C \wedge C\right]=\int d^3x \Tr\left[C\wedge \tilde F 
-\frac{1}{3}C\wedge C\wedge C\right]\cr
&=&\int d^3 x \left [ C^a_i E^a_i 
-\frac13 \epsilon^{ijk}f^{abc}C^a_iC^b_jC^c_k\right]. \label{heldualNA}
\eea

It is straightforward  to realize that this dual singlet  helicity is also  conserved provided that $\vec E^a \cdot\vec B^a=0$, 
true for singlet null fields.

It is worth noting, however, that the dual non-abelian gauge field $C_\mu^a$ is not defined in general (Poincar\'{e} non-Abelian
duality is not defined in general). It is only (in the 
absence of sources, i.e., in vacuum, and) 
in the gauge $C_0^a=0$ that we can define the analog of (\ref{EBa}), replacing $E_i^a\leftrightarrow
B_i^a$, as 
\bea 
E^a_i&=& F^a_{i0}
= \frac12 {\epsilon_{i}}^{jk}(\pa_j C^a_k
-\pa_k C^a_j + {f^a}_{bc} C^b_j C^c_k)\cr
B^a_i&=& \frac12 \epsilon_{ijk} F^{ajk}
=-\d_t C_i^a\;.
\eea

Then in terms of the Poincar\'{e} dual field strengths, 
\be
\tilde F_{\mu\nu}^a\equiv \frac{1}{2}\epsilon_{\mu\nu\rho\sigma}F^{a\rho\sigma}\;,
\ee
we have 
\be
\tilde F^a=dC^a+(C\wedge C)^a.
\ee

Now we can also define non-abelian versions of the mixed abelian helicities ${\cal H}_{me}$ and ${\cal H}_{em}$,  
the helicities ${\cal H}^{NA}_{me}$ and ${\cal H}^{NA}_{em}$.

The natural candidates for these helicities are 
\bea
{\cal H}^{NA}_{em}&=& \int d^3 x \Tr\left[C\wedge dA+ \frac12 A\wedge A \wedge C\right]=\int d^3x \Tr\left[C\wedge F 
-\frac{1}{2}A\wedge A\wedge C\right]\cr
&=&\int d^3 x \left [C^a_i B^a_i 
-\frac12 \epsilon^{ijk}f^{abc}A^a_iA^b_jC^c_k\right]
\eea
and 
\bea
{\cal H}^{NA}_{me}&=& \int d^3 x \Tr\left[A\wedge dC+ \frac12 C\wedge C \wedge A\right]=\int d^3x \Tr\left[A\wedge \tilde F 
-\frac{1}{2}C\wedge C\wedge A\right]\cr
&=&\int d^3 x \left [A^a_i E^a_i 
-\frac12 \epsilon^{ijk}f^{abc}C^a_iC^b_jA^c_k\right]. 
\eea

However, now, for  $\vec{E}^a\vec{E}^a-\vec{B}^a\vec{B}^a=0$, 
true for singlet null fields, these helicities are still not conserved, e.g.,
\be
\d_t {\cal H}_{em}^{NA}=\int d^3x \left[\vec{E}^a\cdot \vec{E}^a-\vec{B}^a\cdot\vec{B}^a+\frac{1}{2}\epsilon^{ijk}f_{abc}
(A_i^aA_j^b B_k^c-C_i^aC_j^bE_k^c)\right]\;,
\ee
by using the YM equation (\ref{YMB}) and partial integration.

But that is not surprising, since having $A_i^a$ and $C_i^a$ defined at the same time is not 
generally consistent: the YM equation for $F^a$ involves $A^a$ explicitly, and is equal to the Bianchi 
identity for $\tilde F^a$, that involves $C^a$ explicitly. But one is defined in the $A_0^a=0$ gauge, and 
the other in the $C_0^a=0$ gauge, and they are not necessarily compatible, unless $A^a\propto C^a$. 
In any case, nonconservation means that we will not use the mixed helicities.


 \subsubsection{ Non-abelian helicities}
 
\noindent 
 
We can also define helicities that are conserved for the non-abelian ($T^a$) null vectors, with 
\be
\Tr[T^a \vec F^2]=0\Rightarrow {d^a}_{bc}\vec{E}^b \cdot\vec{B}^c=0\;\;{\rm and}\;\; 
{d^a}_{bc}\left[\vec{E}^b\cdot\vec{E}^c-\vec{B}^b
\cdot\vec{B}^c\right]=0.
\ee

Indeed, we start with the ${\cal H}_{mm}$ non-abelian conserved helicities, of the form 
\bea
{{\cal H}^{NA}}_{mm}^a &=& \int d^3 x \Tr\left[T^a\left(A\wedge dA+ \frac23 A\wedge A \wedge A\right)\right]\cr
&=&\int d^3x \Tr\left[T^a\left(A\wedge F 
-\frac{1}{3}A\wedge A\wedge A\right)\right]\cr
&=&\int d^3 x\; {d^a}_{bc}\left [A^b_i B^c_i 
-\frac13 \epsilon^{ijk}f^{bde}A^c_iA^d_jA^e_k\right].\label{helNA}
\eea

We now similarly check that for the non-abelian $T^a$-null fields, 
more precisely for the condition ${d^a}_{bc}\vec{E}^b\cdot \vec{B}^c=0$, these non-Abelian helicities indeed are conserved in time.

The same condition leads to the conservation of the helicities
\bea
{{\cal H}^{NA}}_{ee}^a &=& \int d^3 x \Tr\left[T^a\left(C\wedge dC+ \frac23 C\wedge C \wedge C\right)\right]\cr
&=&\int d^3x \Tr\left[T^a\left(C\wedge \tilde F 
-\frac{1}{3}C\wedge C\wedge C\right)\right]\cr
&=&\int d^3 x \;{d^a}_{bc}\left [C^b_i E^c_i 
-\frac13 \epsilon^{ijk}f^{bde}C^c_iC^d_jC^e_k\right].\label{helNA}
\eea

On the other hand, the other half of the $T^a$-null field condition, ${d^a}_{bc}\left(\vec{E}^b\cdot \vec{E}^c-\vec{B}^b\cdot \vec{B}^c
\right)=0$, leads to the {\em almost} conservation of the mixed non-abelian helicities
\bea
{{\cal H}^{NA}}^a_{em}&=& \int d^3 x \Tr\left[T^a\left(C\wedge dA+ \frac12 A\wedge A \wedge C\right)\right]\cr
&=&\int d^3x \Tr\left[T^a\left(C\wedge F 
-\frac{1}{2}A\wedge A\wedge C\right)\right]\cr
&=&\int d^3 x\; {d^a}_{bc}\left [C^b_i B^c_i 
-\frac12 \epsilon^{ijk}f^{bde}A^c_iA^d_jC^e_k\right]
\eea
and 
\bea
{{\cal H}^{NA}}^a_{me}&=& \int d^3 x \Tr\left[T^a\left(A\wedge dC+ \frac12 C\wedge C \wedge A\right)\right]\cr
&=&\int d^3x \Tr\left[T^a\left(A\wedge \tilde F 
-\frac{1}{2}C\wedge C\wedge A\right)\right]\cr
&=&\int d^3 x \;{d^a}_{bc}\left[A^b_i E^c_i 
-\frac12 \epsilon^{ijk}f^{bde}C^c_iC^d_jA^e_k\right]\;,
\eea
except for terms exchanging $A^a_\mu$ with $C^a_\mu$, as in the singlet helicity case.
Again, that means that we will not use them, also since defining $A^a$ and $C^a$ at the same time is 
problematic. 


\section{ Knotted non-abelian solutions}

\noindent

We refer to solutions of the YM equations that carry finite helicity as  knotted non-abelian solutions, 
by analogy to the knotted solutions of the abelian theory like the  
 for instance the solutions found in  \cite{Hoyos:2015bxa,Alves:2017ggb,Alves:2017zjt}.

Knotted solutions of the YM equations fall into three categories:(i) {\it genuine non- abelian knots} which are 
solutions of the YM equations, have finite energy  and are non-abelian, namely, $[A^a,A^b]\neq 0$; (ii) {\it non 
genuine non abelian knots} which solve the YM equations, have finite energy  but  in fact are abelian; and (iii) 
non abelian knots that have divergent energy. 

The process of constructing  knotted solutions that belong to all the three categories   can follow three different  paths: 

(1) Uplifting abelian solutions that carry non trivial helicity, in particular the Hopfion solution and its generalizations, into  solutions of the YM equations. 

(2) Defining non-abelian waves with finite helicity.

(3) Deforming  known non-abelian  solutions that do not carry finite helicities into genuine knotted non-abelian solutions. 

In the next 3 subsections we follow each of the 3  approaches.


\subsection{Uplifting abelian to non-abelian null fields}

In subsection \ref{nullEMsol},  several classes of null abelian solutions were described; in particular the topologically 
non-trivial {\em Hopfion} solution and its generalizations. An obvious question is  
can we find solutions of the YM equations that are singlet (unit) null and/or $T^a$ null? 

The first strategy is to start from  null solutions of Maxwell equations and uplift them to solutions of the YM equations. 
\begin{itemize}
\item
In analogy to the  constant perpendicular abelian electric and magnetic fields one can take the following configuration
\be
A_1^a = z-t\;, \qquad E^a_1= 1\;,\qquad B^a_2=1\;,
\ee
where the gauge fields are non-vanishing only for one particular $a$ and for the rest of the non-abelian indices they vanish.
This is obviously a singlet null non-abelian field configuration, as it is an uplift of the abelian (electromagnetic) case into 
the non-abelian one.

In a similar way, also the plane wave solution
\be
A_1^a =-i e^{i(z-t)}\;,\qquad E^a_1= e^{i(z-t)}\;,\qquad B^a_2=e^{i(z-t)}\;,
\ee
is a null solution of YM equations. 
\item
In fact any solution of the abelian theory can be uplifted to a solution of the non-abelian theory  using the map
\be
A_\mu(t,\vec x)\rightarrow A^a_\mu(t,\vec x)\;,
\ee
for only one particular index $a$. 

Thus the Hopfion solution \cite{Trautman:1977im,Ranada:1989wc,ranada1990knotted,Hoyos:2015bxa} 
and all the infinitely many solutions derived from it are also  null solutions of the YM theory. 
\item
Next we look for null solutions of the YM equations with nonzero fields for more than one $a$.
We first try an ansatz of the form 
\be
A_1^1 = z-t\;, \qquad E^1_1= 1\;,\qquad B^1_2=1\;,
\ee
\be
A_2^2 = z-t\;, \qquad E^2_2= 1\;,\qquad B^2_1=1\;,
\ee
and the rest of the fields zero.

This is obviously null but it does not solve the YM equation for $F^b_{12}$,  for $b=1,2$, so we need to modify the ansatz.

In the gauge $A^a_0=0$ the chromo-electric field is determined by (\ref{EBa}),
namely there is no nonlinear term. The magnetic field does include a nonlinear term of the form $\epsilon^{ijk} f_{abc}A_j^b
A_k^c$, but that vanishes on the above ansatz.

This nonlinear term is also eliminated when the gauge fields have only one nonzero non-abelian index, namely for instance 
\be
A^1_i\neq 0\;,\qquad A^2_i=  A^3_i= 0.
\ee
However, this is in fact the abelian configuration at the item above. 

Another way to eliminate the nonlinear term is if all the components 
$A^a_i$ are equal,  for instance, for the group $SU(2)$ (with $a=1,2,3$),
\be
A_{ai}=\begin{pmatrix}
{\cal A} & {\cal A}  & {\cal A} \\
{\cal A} & {\cal A}  & {\cal A} \\
{\cal A} & {\cal A}  & {\cal A} \\
\end{pmatrix}\;,
\ee
where ${\cal A}= z-t$.
For such a gauge field, we get 
\be
E^a_i=1\;,\qquad B^a_1=-1\;,\qquad B^a_2=1\;,\qquad B^a_3=0.
\ee 

Considering the YM equations for the above, the equation for the electric fields becomes
\be
 (D^i F_{i0})^a =\pa^i F^a_{i0} +\epsilon^{abc}A^i_bF_{i0}^c=0\;,
\ee
because both $A^a_i$ and $E^a_i$ are independent of $a$ and $i$.

For the magnetic fields, the YM equations read
\be
(D^0 F_{0i})^a+(D^j F_{ji})^a =\pa^0 F^a_{0i}+\pa^j F^a_{ji} +\epsilon^{abc} A^i_bF_{ij}^c =0\;,
\ee
so are also satisfied. 

Notice that this configuration obeys
\be
\vec E^a\cdot\vec B^a =0\;,
\ee
but not 
\be
\vec E^a\cdot\vec E^a- \vec B^a\cdot\vec B^a =0
\ee

However, this solution is also abelian, since it can be written as 
\be
A_i^aT_a={\cal A}\left(\sum_a T_a\right)\equiv {\cal A} T\;,
\ee
so we have a single, redefined generator $T$, multiplying all the fields. 

\item

Then, in order to get rid of the non-abelian parts of the YM equations, we can also take a
modified abelian uplift ansatz, where 
\be
A_\mu(t,\vec x)\rightarrow A^a_\mu(t,\vec x)\;,
\ee
but not just for one $a$, rather {\em for all $a$'s}, in particular, in the $SU(2)$ case,
for all $a=1,2,3$.

For instance,  the  constant perpendicular electric and magnetic fields of 
(\ref{Bateconstant}), expressed  in terms of $A_\mu$ as follows:
\be
A_0=-2x\;,\qquad A_x= 2( t+z)\;, \qquad A_y=1\;, \qquad A_z= -2x\;.
\ee

Upon uplifting this to all the three $a=1,2,3$ gauge fields, the YM equations reduce to just the Maxwell ones for each $a$, 
and thus the nontrivial components of the chromo-electric and chromo-magnetic fields become simply
\be
E_x^a= -4\; \qquad B_y^a=4.
\ee
This is just the constant transverse electric and magnetic field solution uplifted to the non-abelian case for all the $a$'s. 

\item
But then, just like it was done in \cite{Hoyos:2015bxa} 
in the case of the electromagnetic solution, we can use a transformation on the 
constant transverse electric and magnetic field solution to obtain the Hopfion solution, with nonzero helicity ${\cal H}_{mm}$.

Thus we can now use the special conformal transformations with imaginary parameter $b_\mu= i(1,0,0,0)$ in this 
nonabelian uplift case, to get the Hopfion solution. The corresponding gauge fields are found to be
\bea
A_0^a&=&\frac{4 t \left(t^4 y-4 t^3 x-2 t^2 y \left(x^2+y^2+z^2+3\right)\right)}{\left(t^4-2 t^2
   \left(x^2+y^2+z^2-1\right)+\left(x^2+y^2+z^2+1\right)^2\right)^2}\CR
   &&+\frac{4 t \left(4 t x \left(x^2+y^2+z^2+1\right)+y
   \left(x^2+y^2+z^2+1\right)^2\right)}{\left(t^4-2 t^2
   \left(x^2+y^2+z^2-1\right)+\left(x^2+y^2+z^2+1\right)^2\right)^2}\;,\CR
A_x^a&=& \frac{4 \left(-2 t^5+t^4 (x y+z)+4 t^3 \left(y^2+z^2\right)-2 t^2 (x y+z) \left(x^2+y^2+z^2+3\right)\right)}{\left(t^4-2 t^2
   \left(x^2+y^2+z^2-1\right)+\left(x^2+y^2+z^2+1\right)^2\right)^2}\CR
   &&+ \frac{4 \left(2 t
   \left(x^2-y^2-z^2+1\right) \left(x^2+y^2+z^2+1\right)+(x y+z) \left(x^2+y^2+z^2+1\right)^2\right)}{\left(t^4-2 t^2
   \left(x^2+y^2+z^2-1\right)+\left(x^2+y^2+z^2+1\right)^2\right)^2}\;,\CR
A_y^a&=&
\frac{2 \left(t^6-t^4 \left(3 x^2+y^2+3 z^2+5\right)+8 t^3 (z-x y)+t^2 \left(3 x^4+2 x^2 \left(y^2+3
   z^2+3\right)-y^4\right)\right)}{\left(t^4-2 t^2
   \left(x^2+y^2+z^2-1\right)+\left(x^2+y^2+z^2+1\right)^2\right)^2}\CR
   &&+
\frac{2\left(t^2 \left(2 \left(y^2+3\right) z^2-6 y^2+3 z^4-5\right)+8 t (x y-z)
   \left(x^2+y^2+z^2+1\right)\right)}{\left(t^4-2 t^2
   \left(x^2+y^2+z^2-1\right)+\left(x^2+y^2+z^2+1\right)^2\right)^2}\CR
   &&+
\frac{2\left(-\left(x^2-y^2+z^2-1\right) \left(x^2+y^2+z^2+1\right)^2\right)}{\left(t^4-2 t^2
   \left(x^2+y^2+z^2-1\right)+\left(x^2+y^2+z^2+1\right)^2\right)^2}\;,\CR
A_z^a&=& \frac{4 \left(2 x^3 \left(-t^2-2 t z+y^2+z^2+1\right)-2 x^2 y \left(-t^2 z+2 t+y^2 z+z^3+z\right)\right)}{\left(t^4-2 t^2 \left(x^2+y^2+z^2-1\right)+\left(x^2+y^2+z^2+1\right)^2\right)^2}\cr
&&+ \frac{4 \left(+x \left(t^4+4 t^3
   z-2 t^2 \left(y^2+z^2+3\right)-4 t z \left(y^2+z^2+1\right)+\left(y^2+z^2+1\right)^2\right)\right)}{\left(t^4-2 t^2 \left(x^2+y^2+z^2-1\right)+\left(x^2+y^2+z^2+1\right)^2\right)^2}\CR
   &&+ \frac{4 \left(-y \left(t^4 z-4 t^3-2
   t^2 z \left(y^2+z^2+3\right)+4 t \left(y^2+z^2+1\right)+z \left(y^2+z^2+1\right)^2\right)+x^5-x^4 y
   z\right)}{\left(t^4-2 t^2 \left(x^2+y^2+z^2-1\right)+\left(x^2+y^2+z^2+1\right)^2\right)^2}\;,\cr
   &&
	\eea
The corresponding components of the electric field  are
\bea
E_x^a &=& \frac{8 \left(t^8-2 t^7 z-2 t^6 \left(2 x^2+y^2+z^2+2\right)+6 t^5 \left(z \left(x^2+y^2+3\right)-2 x y+z^3\right)\right)}{\left(t^4-2 t^2
   \left(x^2+y^2+z^2-1\right)+\left(x^2+y^2+z^2+1\right)^2\right)^3}
\CR
&& +\frac{8 \left(+2
   t^4 \left(3 x^2 \left(x^2+y^2+z^2+2\right)-9 y^2-9 z^2-5\right)+2 t^3 \left(-6 z^3 \left(x^2+y^2+1\right)\right)\right)
   }{\left(t^4-2 t^2
   \left(x^2+y^2+z^2-1\right)+\left(x^2+y^2+z^2+1\right)^2\right)^3}
\CR
&& +\frac{8 \left(+2 t^3 \left(-3 z
   \left(x^2+y^2\right) \left(x^2+y^2+2\right)+4 x y \left(3 x^2+3 y^2+5\right)+12 x y z^2-3 z^5+5 z\right)\right)}
   {\left(t^4-2 t^2
   \left(x^2+y^2+z^2-1\right)+\left(x^2+y^2+z^2+1\right)^2\right)^3}
\CR
&& +\frac{8 \left(-2 t^2
   \left(x^2+y^2+z^2+1\right) \left(2 x^4+x^2 \left(y^2+z^2+4\right)-\left(y^2+z^2\right)^2-11 y^2-11 z^2+2\right)\right)}{\left(t^4-2 t^2
   \left(x^2+y^2+z^2-1\right)+\left(x^2+y^2+z^2+1\right)^2\right)^3}
\CR
&& +\frac{8 \left(+2
   t \left(x^2+y^2+z^2+1\right)^2 \left(z \left(x^2+y^2-5\right)-6 x y+z^3\right)\right)}{\left(t^4-2 t^2
   \left(x^2+y^2+z^2-1\right)+\left(x^2+y^2+z^2+1\right)^2\right)^3}
\CR
&& +\frac{8 \left(+\left(x^2-y^2-z^2+1\right)
   \left(x^2+y^2+z^2+1\right)^3\right)}{\left(t^4-2 t^2
   \left(x^2+y^2+z^2-1\right)+\left(x^2+y^2+z^2+1\right)^2\right)^3}\;,
\CR
E_y^a &=& -\frac{16 \left(2 t^7-t^6 (x y+5 z)+t^5 \left(2-6 y^2\right)+t^4 \left(3 x^3 y+9 x^2 z+3 x y \left(y^2+z^2+5\right)\right)\right)}
{\left(t^4-2 t^2
   \left(x^2+y^2+z^2-1\right)+\left(x^2+y^2+z^2+1\right)^2\right)^3}\CR
&& -\frac{16 \left(+t^4 \left(+9
   z \left(y^2+z^2\right)+5 z\right)-2 t^3 \left(3 x^4+x^2 \left(6 z^2+8\right)-3 y^4-2 y^2+3 z^4+8 z^2+1\right)\right)}{\left(t^4-2 t^2
   \left(x^2+y^2+z^2-1\right)+\left(x^2+y^2+z^2+1\right)^2\right)^3}\CR
   && -\frac{16 \left(-3
   t^2 \left(x^2+y^2+z^2+1\right) \left(z \left(x^2+y^2-3\right)+x y \left(x^2+y^2+5\right)+x y z^2+z^3\right)\right)}{\left(t^4-2 t^2
   \left(x^2+y^2+z^2-1\right)+\left(x^2+y^2+z^2+1\right)^2\right)^3}\CR
   && -\frac{16 \left(+2 t
   \left(x^2+y^2+z^2+1\right)^2 \left(2 x^2-y^2+2 z^2-1\right)+(x y-z)
   \left(x^2+y^2+z^2+1\right)^3\right)}{\left(t^4-2 t^2
   \left(x^2+y^2+z^2-1\right)+\left(x^2+y^2+z^2+1\right)^2\right)^3}\;,\CR		
E_z^a &=& -\frac{16 \left(t^7 x+t^6 (5 y-x z)-3 t^5 \left(x^3+x \left(y^2+z^2+3\right)+2 y z\right)+t^4 \left(-y \left(9 x^2+9
   z^2+5\right)\right)\right)}{\left(t^4-2 t^2
   \left(x^2+y^2+z^2-1\right)+\left(x^2+y^2+z^2+1\right)^2\right)^3}
 \CR
 && -\frac{16 \left(+t^4 \left(+3 x z \left(x^2+z^2+5\right)+3 x y^2 z-9 y^3\right)+t^3 \left(3 x^5+6 x^3 \left(y^2+z^2+1\right)+12
   x^2 y z\right)\right)}{\left(t^4-2 t^2
   \left(x^2+y^2+z^2-1\right)+\left(x^2+y^2+z^2+1\right)^2\right)^3}
 \CR
 && -\frac{16 \left(+t^3 \left(+x \left(3 \left(y^4+2 y^2 \left(z^2+1\right)+z^4\right)+6 z^2-5\right)+4 y z \left(3 y^2+3
   z^2+5\right)\right)\right)}{\left(t^4-2 t^2
   \left(x^2+y^2+z^2-1\right)+\left(x^2+y^2+z^2+1\right)^2\right)^3}
 \CR
 && -\frac{16 \left(+3 t^2 \left(x^2+y^2+z^2+1\right) \left(y \left(x^2+z^2-3\right)-x z \left(x^2+z^2+5\right)-x
   y^2 z+y^3\right)\right)}{\left(t^4-2 t^2
   \left(x^2+y^2+z^2-1\right)+\left(x^2+y^2+z^2+1\right)^2\right)^3}
 \CR
 && -\frac{16 \left(-t \left(x^2+y^2+z^2+1\right)^2 \left(x^3+x \left(y^2+z^2-5\right)+6 y z\right)+(x z+y)
   \left(x^2+y^2+z^2+1\right)^3\right)}{\left(t^4-2 t^2
   \left(x^2+y^2+z^2-1\right)+\left(x^2+y^2+z^2+1\right)^2\right)^3}
 \CR
\eea	
and the components of the magnetic field are:
\bea
B_x^a&=& \frac{16 \left(2 t^7+t^6 (x y-5 z)+t^5 \left(2-6 x^2\right)+t^4 \left(-3 x^3 y+9 x^2 z-3 x y
   \left(y^2+z^2+5\right)\right)\right)}{\left(t^4-2 t^2
   \left(x^2+y^2+z^2-1\right)+\left(x^2+y^2+z^2+1\right)^2\right)^3}
\CR
&& +\frac{16 \left(+t^4 \left(+9 z \left(y^2+z^2\right)+5 z\right)+2 t^3 \left(3 x^4+2 x^2-3 y^4-2 \left(3 y^2+4\right)
   z^2-8 y^2-3 z^4-1\right)\right)}{\left(t^4-2 t^2
   \left(x^2+y^2+z^2-1\right)+\left(x^2+y^2+z^2+1\right)^2\right)^3}
\CR
&& +\frac{16 \left(+3 t^2 \left(x^2+y^2+z^2+1\right) \left(x^3 y-x^2 z+x y \left(y^2+z^2+5\right)-z
   \left(y^2+z^2-3\right)\right)\right)}{\left(t^4-2 t^2
   \left(x^2+y^2+z^2-1\right)+\left(x^2+y^2+z^2+1\right)^2\right)^3}
\CR
&& +\frac{16 \left(-2 t \left(x^2-2 y^2-2 z^2+1\right) \left(x^2+y^2+z^2+1\right)^2-(x y+z)
   \left(x^2+y^2+z^2+1\right)^3\right)}{\left(t^4-2 t^2
   \left(x^2+y^2+z^2-1\right)+\left(x^2+y^2+z^2+1\right)^2\right)^3}\;,
\CR
B_y^a&=& 
\frac{8 \left(t^8-2 t^7 z-2 t^6 \left(x^2+2 y^2+z^2+2\right)+6 t^5 \left(z \left(x^2+y^2+3\right)+2 x y+z^3\right)\right)}{\left(t^4-2 t^2
   \left(x^2+y^2+z^2-1\right)+\left(x^2+y^2+z^2+1\right)^2\right)^3}
\CR
&&+ 
\frac{8 \left(+2
   t^4 \left(3 x^2 \left(y^2-3\right)+3 y^2 \left(y^2+z^2+2\right)-9 z^2-5\right)-2 t^3 \left(6 z^3
   \left(x^2+y^2+1\right)\right)\right)}{\left(t^4-2 t^2
   \left(x^2+y^2+z^2-1\right)+\left(x^2+y^2+z^2+1\right)^2\right)^3}
\CR
&&+ 
\frac{8 \left(-2 t^3 \left(+3 z \left(x^2+y^2\right) \left(x^2+y^2+2\right)+4 x y \left(3 x^2+3 y^2+5\right)+12 x y
   z^2+3 z^5-5 z\right)\right)}{\left(t^4-2 t^2
   \left(x^2+y^2+z^2-1\right)+\left(x^2+y^2+z^2+1\right)^2\right)^3}
\CR
&&+ 
\frac{8 \left(+2 t^2 \left(x^2+y^2+z^2+1\right) \left(x^4+x^2 \left(-y^2+2 z^2+11\right)-\left(y^2-11\right)
   z^2-2 \left(y^2+1\right)^2+z^4\right)\right)}{\left(t^4-2 t^2
   \left(x^2+y^2+z^2-1\right)+\left(x^2+y^2+z^2+1\right)^2\right)^3}
\CR
&&+ 
\frac{8 \left(+2 t \left(x^2+y^2+z^2+1\right)^2 \left(z \left(x^2+y^2-5\right)+6 x
   y+z^3\right)\right)}{\left(t^4-2 t^2
   \left(x^2+y^2+z^2-1\right)+\left(x^2+y^2+z^2+1\right)^2\right)^3}
\CR
&&+ 
\frac{8 \left(-\left(x^2-y^2+z^2-1\right) \left(x^2+y^2+z^2+1\right)^3\right)}{\left(t^4-2 t^2
   \left(x^2+y^2+z^2-1\right)+\left(x^2+y^2+z^2+1\right)^2\right)^3}\;,
\CR
B_z^a&=& \frac{16 \left(-16 t^2 \left(y \left(t^3-t^2 z-t \left(y^2+z^2-1\right)+z \left(y^2+z^2+1\right)\right)+x^2 y (z-t)\right)\right)}{\left(t^4-2 t^2 \left(x^2+y^2+z^2-1\right)+\left(x^2+y^2+z^2+1\right)^2\right)^3}
\CR
&&+ \frac{16 \left(-16 t^2 \left(-x
   \left((t-z)^2+y^2+1\right)-x^3\right)-\left(t^4-2 t^2
   \left(x^2+y^2+z^2-1\right)\right)\right)}{\left(t^4-2 t^2 \left(x^2+y^2+z^2-1\right)+\left(x^2+y^2+z^2+1\right)^2\right)^3}\cr
&&+ \frac{16 \left(-\left(+\left(x^2+y^2+z^2+1\right)^2\right) \left(x \left(5 t^2-6 t z+y^2+z^2+1\right)\right)\right)}{\left(t^4-2 t^2 \left(x^2+y^2+z^2-1\right)+\left(x^2+y^2+z^2+1\right)^2\right)^3}
\CR
&&+ \frac{16 \left(-\left(+\left(x^2+y^2+z^2+1\right)^2\right) \left(+y
   \left(-t^3+t^2 z+t \left(y^2+z^2-5\right)-z \left(y^2+z^2+1\right)\right)\right)\right)}{\left(t^4-2 t^2 \left(x^2+y^2+z^2-1\right)+\left(x^2+y^2+z^2+1\right)^2\right)^3}
\CR
&&+ \frac{16 \left(-\left(+\left(x^2+y^2+z^2+1\right)^2\right) \left(+x^2 y
   (t-z)+x^3\right)\right)}{\left(t^4-2 t^2 \left(x^2+y^2+z^2-1\right)+\left(x^2+y^2+z^2+1\right)^2\right)^3}.
\eea

It is straightforward to check that these are indeed null fields, i.e., 
\be
\vec E^a\cdot\vec B^a=0\;,\qquad \vec E^a\cdot\vec E^a-\vec B^a\cdot\vec B^a=0.
\ee

Using the values of $\vec A^a$ and $\vec B^a$, we find that the magnetic singlet non-abelian helicity is
\bea
{\cal H}_{mm}^{NA}&=& \int d^3 r \vec A^a\cdot \vec B^a\cr
&=& 3\int d^3 r \frac{16 \left((t-z)^2+x^2+y^2+1\right)}{\left(t^4-2 t^2
   \left(x^2+y^2+z^2-1\right)+\left(x^2+y^2+z^2+1\right)^2\right)^2}=
	\frac{3\pi^2}{16}.\cr
	&&
\eea

\item
It was shown in \cite{Kedia:2013bw} that from any Bateman solution $(\alpha,\beta)$ one can generate families of solutions using the following transformation
\be
(\alpha,\beta) \rightarrow (g((\alpha,\beta),h(\alpha,\beta))
\ee
where $g$ and $f$ are any holomorphic functions. In particular a set of infinitely many solution are generated with $(f,g)= (\alpha^p,\beta^q)$. Fig (\ref{ep2q3}) illustrates  the $(p=2,q=3)$ knot solution described in terms of the isosurface $\phi_E$ defined by
$\vec E\cdot\vec \nabla \phi_E=0$.
\begin{figure}[ht!]
			\centering
				\includegraphics[ width=0.85\textwidth]{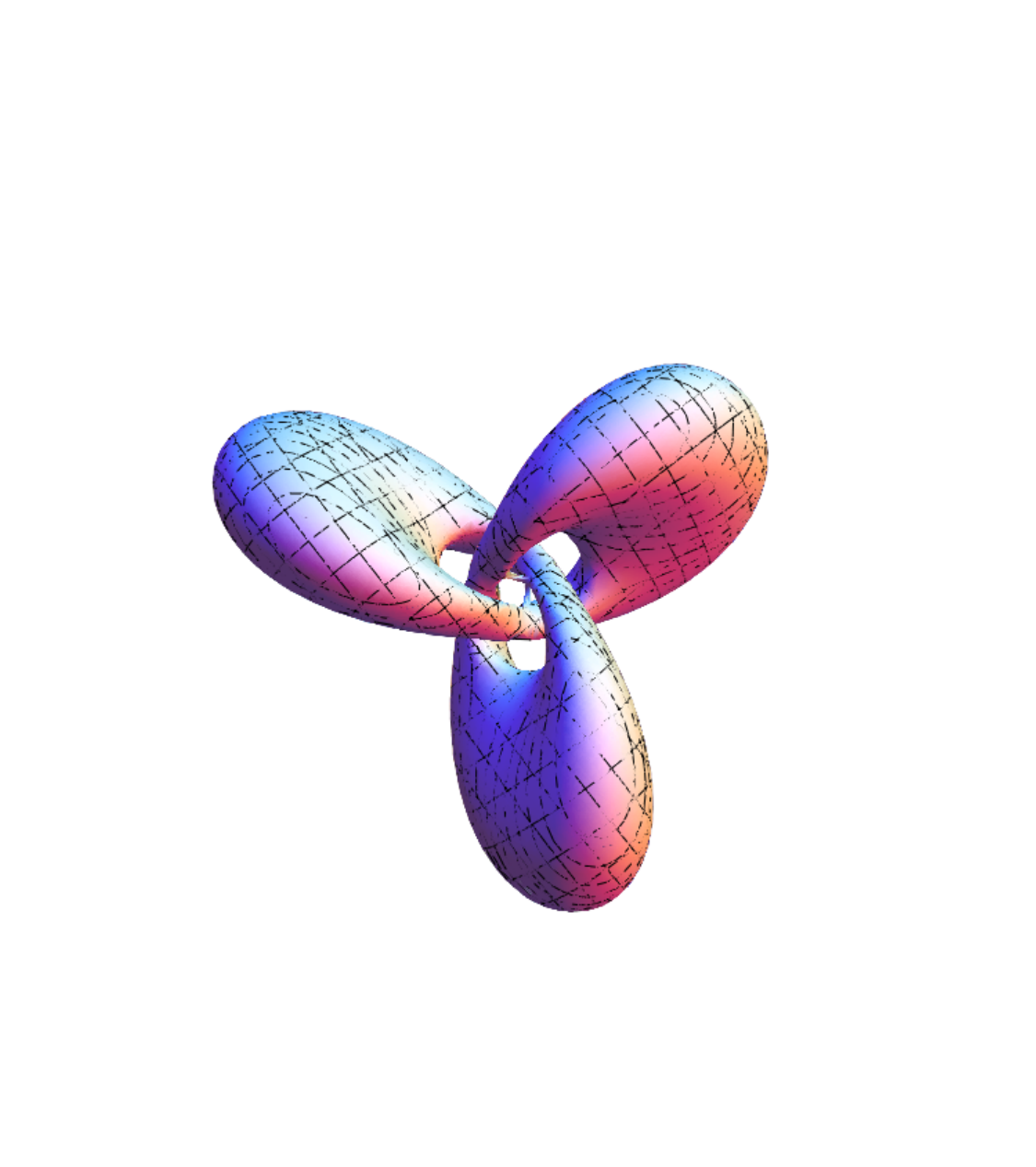}
				\caption{\label{ep2q3}Isosurface with $\phi_E=0.05$ of the  (2,3) knot solution at $t=0$. \cite{Hoyos:2015bxa}}
	\end{figure}

\end{itemize}



\subsection{Radial non-abelian wave}

\noindent

Following the ansatz for a radial nonabelian wave in \cite{SingaporeLi}, we write 
\be
A_\mu^a=\sum_{\a=1}^4\psi_\a f_\a ^a(r\pm t)\d_\mu (r\pm t)\;,
\ee
in which case the YM equations reduce to 
\bea
&&(\d_r\mp \d_t)\psi_\a=0\cr
&&\d_\nu(r\pm t)\Box \psi_\a+\d_\nu(\d_r\mp\d_t)\psi_\a=0.
\eea

Then, $\psi_\a=\psi_\a(\theta,\phi)$ means the equations reduce to  just $\Box \psi_\a(\theta,\phi)=0$, which can be solved 
by 
\be
\psi_1=\phi\;,\;\;
\psi_2=\ln\left(\tan\frac{\theta}{2}\right)\;,\;\;
\psi_3=\psi_4=0.
\ee

Note that $f_\a^a(r\pm t)$ are unrestricted functions, which we will fix by the condition to obtain useful solutions with finite 
helicity.

Taking the upper sign, for functions of $r+t$, we obtain the gauge field 
\be
A_\mu^a=-\phi f_1^a\d_\mu(r+t)+\ln \left(\tan \frac{\theta}{2}\right)f_2^a\d_\mu(r+t)\;,
\ee
so in components
\be
A_r^a=A_0^a=\phi f_1^a+\ln\left(\tan\frac{\theta}{2}\right)f_2^a\;,\;\;\; A_\theta^a=A_\phi^a=0\;,
\ee
and the chromo-electric and chromo-magnetic fields 
\bea
&& E_r^a=0\;,\;\; E_\theta^a=-\frac{f_2^a}{\sin\theta}\;,\;\; E_\phi^a=-f_1^a\cr
&& B_r^a=0\;,\;\; B_\theta^a=\frac{f_1^a}{\sin\theta}\;,\;\; B_\phi^a=-f_2^a.
\eea

Then 
\be
\vec{E}^a\cdot \vec{B}^a=0=\vec{E}^a\cdot\vec{E}^a-\vec{B}^a\cdot\vec{B}^a=0\;,
\ee
i.e., we have singlet null fields, so the singlet non-abelian helicities should be conserved in time. 

Moreover, 
\be
\vec{A}^a\cdot \vec{B}^a=g^{rr}A_r^aB_r^a=\frac{f_1^a}{\sin\theta}\left[\phi f_1^a+\ln\left(\tan\frac{\theta}{2}\right)f_2^a\right]\;,
\label{ABnonab}
\ee
so the $mm$ singlet non-abelian helicity is 
\bea
{\cal H}_{mm}^{NA}&=& \int _0^\infty dr\; r^2\int_0^{2\pi}d\phi \int_0^\pi d\theta\sin\theta\vec{A}^a\cdot \vec{B}^a\cr
&=& \int_0^\infty dr\; r^2\left[(f_1^a)^2\frac{(2\pi)^2}{2}+(f_1^a f_2^a)2\pi 2\int_0^{\pi}d(\theta/2)\ln \tan(\theta/2)\right]
\eea
and since the second integral vanishes, we finally get
\be
{\cal H}^{NA}_{mm}=2\pi^2\int_0^\infty dr\; r^2(f_1^a)^2\;,
\ee
which can be easily made finite, as well as intrinsically non-abelian. 

Note that the helicity density in (\ref{ABnonab}) is discontinuous at $\phi=0\sim 2\pi$, when $\phi\in [0,2\pi)$, making the 
choice of origin for $\phi$ special in this case.

Intrinsically non-abelian means that that for each of the generators $T_a$, $a=1,2,3$ (for the group $SU(2)$), 
we have a different function of $r+t$. We must have a finite ${\cal H}_{mm}^{NA}$, as well as functions that are finite at $r=0$
and actually, at any $r$. The simplest such choice for $f_1^a$ (appearing in ${\cal H}_{mm}^{NA}$) is 
\be
f_1^1=\frac{a}{(r+t)^4+a^4}\;,\;\;
f_1^2=\frac{a^3}{(r+t)^6+a^6}\;,\;\;
f_1^3=\frac{a^5}{(r+t)^8+a^8}. 
\ee

Then $A_\mu^a\sim f_1^a\sim 1/r^4$ at infinity, and generically $E_i^a, B_i^a$ are also $\sim f_1^a\sim 1/r^4$, however 
$E_\theta^a, B_\theta^a$ are singular at $\theta=0,\pi$. So, {\em if we exclude a small tube around $\theta=0,\pi$}, 
the space is effectively compactified, and we would have a Hopfion. Of course, the singularity at $\theta=0,\pi$,
besides the discontinuity of the helicity density at $\phi=0$, precludes 
such interpretation,  even though ${\cal H}_{mm}^{NA}$ is finite. 

But we still need to see whether the energy is finite. We have
\bea
{\cal E}&=& \int_0^\infty r^2\; dr \int_0^{2\pi}d\phi\int_0^{\pi}\sin \theta d\theta\left[E_\theta^aE_\theta^ag^{\theta\theta}
+E_\phi^aE_\phi^a g^{\phi\phi}\right]\cr
&=& 2\pi \int_0^{\pi}\frac{d\theta}{\sin\theta}\int_0^\infty dr \left(f_1^af_1^a+f_2^af_2^a\right).
\eea

The integral over $r$ can easily be made finite (in fact, it is finite for the $f_1^a$ defined above, and we can 
put also $f_2^a=f_1^a$, for instance). However, the integral over $\theta$ is divergent: regularizing it, we get
\be
-\ln\left|\frac{1+\cos x}{\sin x}\right|_\epsilon^{\pi-\epsilon}=-\left(\ln \frac{\epsilon}{2}-\ln \frac{2}{\epsilon}\right)=
-2\ln \frac{\epsilon}{2}.
\ee

So, unfortunately, we cannot interpret this solution as a finite energy knotted solution.    It is   an infinite energy wave.

\subsection{Deforming trivial solutions to ones that admit finite helicities}

\noindent

Other kinds of solutions can be obtained by noting, as we did, that ${\cal H}_{mm}^{NA}$ is not gauge invariant, if we 
take gauge transformations that have nontrivial values at the boundaries of space. 

\subsubsection{Non-abelian plane wave without helicity}

\noindent

One such solution is a nonabelian plane wave.

We start by considering the plane wave ansatz (for group $SU(2)$), also from \cite{SingaporeLi},
\be
A_0^a=-A_3^a=-f_\a ^a(z+t) g_\a (x,y)\;,\;\;\; A_1^a=A_2^a=0\;,
\ee
in which case 
we find that the YM equations reduce to 
\be
f_\a ^a(\d_x^2+\d_y^2)g_\a (x,y)=0.
\ee

Moreover, then one finds 
\bea
&& E_3^a=2g_\a {f'_\a} ^a\;,\;\;
E_1^a=+f^a_\a \d_x g_\a\;,\;\;
E_2^a=+f^a_\a \d_y g_\a\cr
&& B_3^a=0\;,\;\; B_1^a=+f^a_\a \d_y g_\a\;,\;\;
B_2^a=-f^a_\a \d_x g_\a\;,
\eea
so that 
\be
\vec{E}^a\cdot\vec{E}^a-\vec{B}^a\cdot \vec{B}^a=0\;,\;\; \vec{E}_a\cdot\vec{B}_a=0\;,
\ee
meaning again we have singlet null nonabelian fields, but also 
\be
A_i^aB_i^a=0\;,\;\; \epsilon^{ijk}\epsilon^{abc}A_i^aA_j^b A_k^c\propto \epsilon^{333}=0\Rightarrow 
H_{mm}^{NA}=0\;,
\ee
so unfortunately we have no singlet  helicity. 

But, in order to get something nontrivial, we can make a gauge transformation on $A_i^a$ 
that is nontrivial at the boundaries of space (so that $B_i^a$ is unchanged, but 
the helicity is changed). 

We first choose 
\be
f_\a^a(z+t)=(z+t)\delta_\a^a\;,\;\;\; g_1=x^2-y^2;,\;\; g_2=x-y\;,\;\; g_3=0.
\ee

This is a truly non-Abelian solution, in that we get (since $T_a=\sigma_a$, the Pauli matrices)
\be
A_0^a=-A_3^a=-(z+t)(x-y)[(x+y)\sigma^1+\sigma^2]\equiv -(z+t)(x-y)\sigma\;,
\ee
but $\sigma=(x+y)\sigma^1+\sigma^2$ is not a constant matrix in the Lie algebra (which would result in 
an Abelian solution), but rather is an $(x,y)$-dependent matrix (so that matrices at different positions $(x,y)$ don't commute). 

Moreover then, choose a gauge transformation with 
\be
U=e^\sigma\Rightarrow U^{-1}A_\mu U =A_\mu\;,\;\; U^{-1}\d_\mu U =\d_\mu \sigma\;,
\ee
so that the transformed gauge field is 
\be
A'_\mu =A_\mu +\d_\mu \sigma\Rightarrow A'_1=A_1+\sigma_1=\sigma_1\;,\;\; 
A'_2 =A_2 +\sigma^1=\sigma^1\;,\;\; A'_3=A_3.
\ee

Then 
\be
\vec{A}'_a\cdot\vec{B}_a=-2(x+y)(z+t)\;,\;\;
\epsilon^{ijk}\epsilon_{abc}{A'_i}^a{A'_j}^b {A'_k}^c=0\;,
\ee
so 
\be
H_{mm}=-2\int dx\; dy\; dz (x+y)(z+t).
\ee

But this is naively divergent, so it is hard to interpret. 
Moreover, using the simplest regularization, with $\int_{-\Lambda}^{+\Lambda}dx
\int_{-\Lambda}^{+\Lambda}dy\; (x+y)$, we get zero again!

\subsubsection{Nonabelian plane wave solution with finite helicity}

\noindent

But we note that, in fact, we can choose 
\be
g_1=x^2-y^2+a(x-y)\;,\;\; g_2=c(x-y)\;, g_3=0\;,
\ee
with $a,c$ constants with dimensions of length,
which means $\sigma=\frac{1}{a}[(x+y+a)\sigma^1+c\sigma^2]$ (note that we could multiply this in principle with any 
-dimensionless- number), and then use it in $U=e^\sigma$, which means again 
\be
A'_\mu =A_\mu +\d_\mu \sigma\Rightarrow A'_1=A_1+\frac{1}{a}\sigma_1=\frac{1}{a}\sigma_1\;,\;\; 
A'_2 =A_2 +\frac{1}{a}\sigma_1=\frac{1}{a}\sigma_1\;,\;\; A'_3=A_3.
\ee

Then 
\be
\vec{A}'_a\cdot\vec{B}_a=-\frac{1}{a}(2x+2y+a)f(z+t)\;,
\ee
since we note also that {\em any} ansatz 
\be
f^a_\a=\delta_\a ^a f(z+t)
\ee
works as well, the only change is that now
\be
B_3^a=0\;,\;\; B_1^a=f(z+t) \d_y g^a\;,\;\; B_2^a=f(z+t) \d_x g^a\;.
\ee

It only remains to choose an appropriate $f(z+t)$ such that, at fixed $t$, regularizing the integrals by $\int_{-\Lambda }^{+\Lambda}
$ in all $x,y,z$, we get a finite result, and moreover we don't introduce a singularity at 0. Since 
\be
\int_{-\Lambda}^{+\Lambda } dx \int _{-\Lambda}^{+\Lambda} dy \; (2x+2y+a)=4a\Lambda^2\;,
\ee
we need something that goes like $1/\Lambda^2$ for the integral $\int_{-\Lambda}^{+\Lambda}dz f(z+t)$.

This case is obtained, for instance, for 
\be
f(z+t)=\frac{1}{a_1^2-a_2^2}\left[\frac{a_1^2|z+t|}{(z+t)^4+a_1^4}-\frac{a_2^2|z+t|}{(z+t)^4+a_2^4}\right].
\ee

Note that the ansatz is good, since we still have $(\d_z^2-\d_t^2) A_3^a=0$, since $(\d_z^2-\d_t^2)|z+t|\sim \delta(z+t)
-\delta(z+t)=0$. 

Note that this ansatz is nonsingular everywhere, and {\em if all the spatial coordinates scale similarly},
$f\sim 1/r^3$, so $A_\mu\sim 1/r$ 
+ a constant at infinity, so space would be effectively compactified to $S^3$, and we would now obtain $B\sim 1/r^2$. 
But, as in the case of the radial wave, there are special directions in the space 
(namely, $z+t$ fixed, but $x,y$ go to infinity) along which the gauge field and the 
magnetic field do not go to zero at infinity, meaning the compactified space still misses a direction. 

The helicity is finite, since 
\be
\int dx \frac{x}{1+x^4}=\frac{1}{2}{\rm arctan}\; x^2\Rightarrow \int_{-\Lambda/b}^{+\Lambda/b}dx\frac{|x|}{1+x^4}
=\frac{\pi}{2}-\frac{b^2}{\Lambda^2}+\frac{b^6}{2\Lambda^6}+...\;,
\ee
it follows that (at fixed $t$, and $\Lambda \gg t$)
\be
\int_{-\Lambda}^{+\Lambda}dz f(z+t)\simeq \frac{1}{a_1^2-a_2^2}\left[-\frac{a_1^2}{\Lambda^2}+\frac{a_2^2}{\Lambda^2}\right]
=-\frac{1}{\Lambda^2}\;,
\ee
such that the helicity is 
\be
{\cal H}^{NA}_{mm}=4.
\ee

The energy is, however, still infinite, just like in the case of the radial wave, as we can easily check, since as before, the 
integral over $x,y$ still diverges, but the integral over $z$ is now finite, so the full result diverges.

\subsubsection{Abelian monopole}

\noindent

To better understand these cases, of the radial and plane waves with finite helicity, we go back to the abelian case, and 
consider the abelian monopole configuration.

For the usual Dirac monopole, 
\be
\vec{B}=\frac{\tilde g}{r^2}\hat r\;,
\ee
with $\tilde g=\mu_0 g/(4\pi)$, $\hat r=\vec{e}_r=\vec{r}/r$,
the usual gauge field, in the 2 patches ($\a$ excludes the South Pole, $\b$ excludes the North Pole) is 
\be
\vec{A}^{(\a)}=\frac{\tilde g}{r}\frac{(-1+\cos\theta)}{\sin \theta}\vec{e}_\phi\;,\;\;
\vec{A}^{(\b)}=\frac{\tilde g}{r}\frac{(+1+\cos\theta)}{\sin\theta}\vec{e}_\phi\;,
\ee
so that $\vec{A}\cdot\vec{B}=0$, so that this monopole has ${\cal H}_{mm}=0$. 

However, we can also make a gauge transformation that takes ${\cal H}_{mm}$ to a nonzero value, just like in the 
case of the nonabelian plane wave above. 

For instance, take a gauge transformation with parameter
\be
\lambda=\frac{q}{\sqrt{r^2+a^2}}=\frac{q/a}{\sqrt{(r/a)^2+1}}\;,
\ee
so $\vec{A}'=\vec{A}+\vec{\nabla}\lambda$, and then
\be
\vec{A}'=\vec{A}+\frac{r}{(r^2+a^2)^{3/2}}\hat r\;,
\ee
which gives
\be
\vec{A}'\cdot \vec{B}=\frac{q\tilde g }{r(r^2+a^2)^{3/2}}.
\ee

Then 
\bea
{\cal H}_{mm}&=& \int 4\pi r^2dr\vec{A}'\cdot \vec{B}=4\pi q\tilde g \int \frac{rdr}{(r^2+a^2)^{3/2}}
=-\left.\frac{4\pi q\tilde q}{(r^2+a^2)^{1/2}}
\right|_a^\infty\cr
&=& \frac{4\pi q\tilde g}{a}\;,
\eea
and we see that $a\rightarrow 0$ takes ${\cal H}_{mm}\rightarrow \infty$. 

This result makes sense, since the gauge variation of the helicity, as a CS term, is 
\bea
\delta \int_V  A\wedge d A&=&\int_V  F\wedge d\lambda=\int_V d(F\lambda)=\oint_{\rm bd.} F\lambda\cr
&=& \left.4\pi r^2\vec{B}\lambda\right|_{r=0}^{r=\infty}=\frac{4\pi q\tilde g}{a}\;,
\eea
as before (since the variation of the helicity, as a  CS term, is from 0 to the nonzero value). 
We see that the boundary at infinity has no contribution, i.e., $\lambda$ goes to zero sufficiently fast at infinity. 
However, the nontrivial contribution is from $r=0$, where $\lambda $ is finite, but $\vec{B}$ diverges. 

Note then that ${\cal H}_{mm}$ has now a {\em finite} value, though again this cannot be reinterpreted in terms of a 
nonzero Hopf index, since the reason we obtained ${\cal H}_{mm}\neq 0$ is because we have excluded $r=0$ from the 
space (it is a "boundary"), thus changing the topology from $S^3$ back to $\mathbb{R}^3$ (we gained the point at infinity, 
which would compactify space, but we lost the point at zero, so we are back to decompactified).

\subsubsection{'t Hooft-Polyakov monopole and finite helicity "Hopfion"}

\noindent

But we can use the example of the Dirac monopole to find a nonabelian 
solution with finite singlet nonabelian helicity that is nonsingular, at least. 

This example, somewhat in between the Dirac monopole and the Hopfion, starts with  the 't Hooft-Polyakov monopole, 
with the standard ansatz
\bea
\phi^a&=& \frac{x^a}{r}\phi_0 h(\phi_0 gr)\;,\cr
A_i^a&=& -\epsilon_{ija}\frac{x^j}{gr^2}[1-K(\phi_0 gr)]\;,
\eea
which for the {\em BPS saturated} monopole becomes 
\bea
E_i^a&=& 0\;,\;\; B_i^a=(D_i \phi)^a\;,\;\; (D_0\phi)^a=0\cr
h(r)&=& \frac{1}{\tanh 2r}-\frac{1}{2r}\cr
K(r)&=& \frac{2r}{\sinh 2r}.
\eea

Then we obtain 
\bea
B_i^a&=& \d_i \phi^a+[A_i,\phi]^a=\delta_a^i\frac{\phi_0}{r}h(\phi_0 gr)+\frac{x^ax^i}{r}\phi_0\left[\frac{h(\phi_0 gr)}{r}\right]'\cr
&&+(\delta_i^ar^2-x^i x^a)\frac{1-K(\phi_0 gr)}{gr^2}\frac{\phi_0 h(\phi_0 gr)}{r}\;,
\eea
and we see that: it is of order $1/r^2$ at $r\rightarrow \infty$, and has only $x^i x^a$ and $\delta_i^a$ terms (both symmetric). 
Thus when contracted with $A_i^a\propto \epsilon_{ija}$, we get zero. We also get zero 
from $\epsilon^{ijk}\epsilon_{abc}A_i^aA_j^b A_k^c$, since we get 3 $x$'s contracted, and there is no symmetric Lorentz invariant 
with 3 indices. Thus we get zero singlet nonabelian helicity ${\cal H}^{NA}_{mm}$. 

However, again we can go to a different gauge, by a gauge transformation with parameter
\be
U=\exp \left[i\tanh(\phi_0r)(\phi^a-\phi^a_{\rm vac})T_a/\phi_0\right]\equiv e^{if(r)(x^aT_a)}\;,
\ee
where $\phi_{\rm vac}^a=\phi_0 x^a/r$ and 
$f(r)\sim 1/r$ at $r\rightarrow \infty$. Note that, because of the choice of $U$, the scalar field $\phi^a$ 
doesn't change 
its boundary conditions.

Then we have ${A'_i}^a=U^{-1}A_i U +U^{-1}\d_i U$, and
\bea
U^{-1}\d_i U& =& T_a \left[\delta_i^a+\frac{x^ax_i }{r^2}\right]\times ({\rm fct.of\;r})\sim 1/r^2\cr
U^{-1}A_i U&\simeq & A_i +[A_i , f(r)(x^aT_a)]+...\cr
& \simeq& A_i +\left(\delta_i^a-\frac{x^ax_i}{r^2}\right)T_a\times ({\rm fct. of\; r})\sim A_i +{\cal O}(1/r^2)\;,
\eea
which means that, as for the Dirac monopole, the singlet nonabelian helicity ${\cal H}^{NA}_{mm}$ is 
now nonzero and finite at infinity 
(since at infinity, ${A'}_i^aB_i^a\sim \frac{1}{r^4}$ and, when integrated over $r^2dr$, gives a finite result). 
We were not able to find any modified helicity that is invariant under this gauge transformation, though we cannot 
exclude the possibility.

However, unlike the Dirac monopole (for the case we take $a=0$, 
such that the solution is only defined by a $1/r$ function), 
it is also finite at $r=0$. 
Note that the gauge transformation $U$, and the transformed gauge field $A'_i$, are 
finite and single-valued at $r=0$, because the same is true for $A_i^a$ and $r(\phi^a-\phi^a_{\rm vac})$ 
(so $U\simeq 1$ and $U^{-1}\d_i U\simeq \phi_0\delta_i^aT_a$ at $r\simeq 0$).
That means that the helicity is now defined by a field configuration that is truly finite. 
And we know that the field strengths of the 't Hooft monopole are such that the energy is finite also, unlike the 
wave and Dirac monopole cases before. 

So unlike the Dirac monopole case, now {\em the solution with helicity is 
defined by the scalar function $\phi^a$}, and over the whole 
$S^3=\mathbb{R}^3\cup \{\infty\}$, so it  can be interpreted as a sort of {\em nonabelian Hopfion}. 

\subsection{Proposals for a nonabelian generalization of the Bateman formulation}

\noindent

For a non-abelian group ${\cal G}$, with a non-abelian algebra  $G$, we take two group elements 
\be
{\cal A}\in {\cal G}\Rightarrow {\cal A}^{-1}\pa_\mu {\cal A}\in  G \;, \qquad {\cal B}\in {\cal G}\Rightarrow {\cal B}^{-1}\pa_\mu {\cal B}\in  G.
\ee

For the case of an $SU(2)$ group, with generators $T_a=\sigma_a$, $a=1,2,3$,
we use the well-known parametrization of the $2\times 2$ matrices 
in terms of $\sigma_\a=(\one,\sigma_a)$ to define $2\times 2$ matrices with the extra component (0) defined through 
normalization,
\bea
{\cal A} &=& {\cal A}^\mu\sigma_\mu\;, \qquad  |{\cal A}^0|^2 +\sum_{a=1}^3 |{\cal A}^a|^2 =1\;, \CR
 {\cal B}&=& {\cal B}^\b\sigma_\b\;, \qquad  |{\cal B}^0|^2 +\sum_{a=1}^3 |{\cal B}^a|^2 =1.
\eea

There are several ways to attempt a generalization of the Bateman construction for non-abelian gauge fields. 

1) One possible approach is to start with the gauge fields themselves and write a Bateman-like formula for them. 
Define the complex non-abelian gauge field
\be
H_\mu= H^a_\mu T^a = C_\mu+ i A_\mu\;, \qquad H^a_\mu = C^a_\mu+ i A^a_\mu\;,\;\; \mu=(0,i).\label{Bgaugefields}
\ee

Assume the following form for $H_\mu$ in terms of ${\cal A},{\cal B}\in\mathbb{C}$ (generalizing $\a,\b\in\mathbb{C}$ for the 
abelian case, for the imaginary part = $A_\mu$):
\be
H_\mu = \frac 12 [{\cal A}\;\pa_\mu {\cal B}-\pa_\mu{\cal A}\; {\cal B}].
\ee

The corresponding  two-form field strength tensor $K$ for $H$  is assumed to be (such that the regular field strength $F$
for $A$ is part of $K$, but note that this implies some unusual $-i$'s) 
\be
K = G+iF\;, \qquad K = dH -i H\wedge H\;,  \qquad G = dC -i C\wedge C\;, \qquad F = dA + A\wedge A\;,
\ee
so that in (complex) components
\be  
K_{\mu\nu}= \pa_\mu{\cal A}\pa_\nu{\cal B}
-\pa_\nu{\cal A}\pa_\mu{\cal B}-i({\cal A}\pa_\mu {\cal B}
-\pa_\mu{\cal A} {\cal B})({\cal A}\pa_\nu {\cal B}-\pa_\nu{\cal A} {\cal B}).
\ee

Note that  if we write $C=i\tilde C$, $H=i\tilde H$, $G=i\tilde G$,  (purely imaginary fields) we have the usual $\tilde G=d\tilde C
+\tilde C\wedge \tilde C$, $\tilde K=d\tilde H+\tilde H\wedge \tilde H$, but then $K=i\tilde K$ is purely imaginary, 
and $\tilde K=\tilde G+F, \tilde H=\tilde C+A$ are real, which is not what we want.
 
We could continue with this $K_{\mu\nu}$, since it contains already half the YM equations (as $K$ is written as $dH-iH\wedge H$),
and find the what is the condition for the other half ($D^\mu F_{\mu\nu}=0$) to be satisfied, in terms of $F$ and $G$. 

2) But another proposal is to generalize the Bateman formulation directly for the Riemann-Silberstein vector, to
\be
\vec F\equiv \vec E+i\vec B 
=  ({\cal A}^{-1}\vec D {\cal A})\times ({\cal B}^{-1}\vec D  {\cal B})\Rightarrow F^i=\epsilon^{ijk}({\cal A}^{-1}D_j{\cal A})
({\cal B}^{-1}D_k{\cal B})\;,\label{nonabF1}
\ee
 where 
\be
 D {\cal A}=d {\cal A} + [ A ,{\cal A}]
\ee
 is the covariant derivative 
which includes also the non-abelian  gauge connection. However, because of the term with $A$, we see that we need to 
have also a Bateman formula for $A$ itself. 

We can also generalize these formulas to full $2\times 2$ matrices, using the $\sigma^\a$ matrices, as above. 
In terms of ${\cal A}={\cal A}^\a\sigma_\a $,  the action of the covariant derivative reads
\be
\vec D  {\cal A}^0= \vec \nabla {\cal A}^0\;, \qquad
\vec D  {\cal A}^a= \vec \nabla {\cal A}^a + g{f^a}_{bc} \vec A^b{\cal A}^c\;,
\ee
and similarly for ${\cal B}$, so on the zero component, the covariant derivative acts trivially. However, in the following, 
we will continue with only $SU(2)$ matrices.

Next we can insert this ansatz into the YM equations of motion, which in terms of $\vec{F}=\vec {E}+i\vec{B}$ become
\be
\vec{D}\cdot\vec{F}=0\;;\;\;\;
\d_t\vec{F}+i\vec{D}\times \vec{F}=0\;.\label{YMF}
\ee

Thus, we need also an expression for the gauge field in terms of the generalized Bateman variables ${\cal A}$ and 
${\cal B}$. 
In analogy to the {\em on-shell} abelian gauge field,
\be
A_\mu= {\rm Im}[ \alpha\pa_\mu \beta- \beta\pa_\mu \alpha]\;,
\ee
and having in mind consistency with (\ref{nonabF1}) (which is true, since for small $A_\mu^a$, meaning small ${\cal A}, 
{\cal B}$, we get the correct field strength in (\ref{nonabF1}), and $A_\mu^a$ is added to give the covariant derivative $D$
instead of $d$ in a unique way)
we propose the {\em on-shell} expression
\be
A^a_\mu= {\rm Im}[ -i (\log[{\cal A}] {\cal B}^{-1}\pa_\mu {\cal B}- \log[{\cal B}] {\cal A}^{-1}\pa_\mu {\cal A})]^a.\label{Anonab1}
\ee

With this ansatz, the YM equations take the form
\bea
&&\vec \nabla\cdot \vec F^a + g{f^a}_{bc} {\rm Im}[ -i (\log[{\cal A}] {\cal B}^{-1}\vec{\nabla} {\cal B}- \log[{\cal B}] {\cal A}^{-1}\vec
{\nabla} 
{\cal A})]^b \cdot\vec{F}^c = 0\cr
&& \d_t\vec{F}^a+i g{f^a}_{bc} {\rm Im}[ -i (\log[{\cal A}] {\cal B}^{-1}\vec{\nabla} {\cal B}- \log[{\cal B}] {\cal A}^{-1}\vec
{\nabla} 
{\cal A})]^b \times\vec{F}^c=0\;,
\eea
and we need to substitute (\ref{nonabF1}) in it, and see under what conditions it is satisfied. In fact, like in the abelian case,
the first one is automatically satisfied by (\ref{nonabF1}), and the second is satisfied by considering also the $\mu=0$ 
component of (\ref{Anonab1}).


\subsection{ Examples of non-abelian Bateman solutions}

Using the first approach of expressing the gauge fields in terms of two complex group elements ${\cal A},{\cal B}\in 
\mathbb{C}$ (\ref{Bgaugefields}), 
we take the following ansatz for the Bateman fields
\bea
&&\alpha^1= 2i(z+t) -1 \qquad \beta^1 = 2(x-iy)\;,\qquad  \alpha^2 = 2(x-iy)\;,\qquad  \beta^2= 2i(z+t) -1 \cr
&&\alpha^0=\alpha^3=\beta^0= \beta^3=0\;,
\eea
namely, we swapped the roles of $\alpha$ and $\beta$ for the non-abelian indices $a=1$ and $a=2$.

Using (\ref{Bgaugefields}) we get the following gauge fields
\bea
A^1_0 &=& -2x\;, \qquad  A^1_x =  2(z+t)\;,\qquad  A^1_y =  1\;,\qquad  A^1_z=-2x \cr
A^2_0 &=& 2x\;, \qquad  A^2_x = -2(z+t)\;, \qquad  A^2_y =-1\;, \qquad  A^2_z=2x.\label{nonabBate}
\eea

The only non-abelian field strengths are
\be
F^1_{10}=-4\;, \qquad F^1_{31}= 4\;,\qquad  F^2_{10}=4\;, \qquad F^2_{31}= -4.
\ee  

Notice that terms of $F^3_{\mu\nu}$  that could have emerged from the term $g f^3_{12} A^1_\mu A^2_\nu $ turn out to vanish.

However, also notice that the gauge fields in (\ref{nonabBate}) are obtained by replacing the generators $T^1$ with 
${T' }^ 1=T^ 1-T^ 2$, so this solution is in fact abelian, and equivalent to the solution in subsection 4.1 


The next step is transforming $\alpha^\mu,\beta^\mu$ using the special conformal transformation 
characterized by  $b_\mu= i(1,0,0,0)$ to get 
\be
\a^1=\b^2=\frac{A-iz}{A+it}\;,\;\;
\b^1=\a^2=\frac{x-iy}{A+it}\;,\;\;
A=\frac{1}{2}(x^2+y^2+z^2-t^2+1).\label{nabsol}
\ee

In this way we get non-abelian gauge fields $A^1_\mu$ and $A^2_\mu$ that are Hopfion configuration with ${\cal H}_{mm}=1$.

\section{Conclusions and open questions}

In this paper we have generalized the notions of null fields, helicity and knotted solutions to the case of Yang-Mills theory. 
We have first defined singlet-null fields as $\Tr[\vec{F}^2]=0$ and $T^a$-null fields as $\Tr[T^a\vec{F}^2]=0$. Then 
we have defined singlet helicities as the singlet spatial CS forms, for instance ${\cal H}_{mm}^{NA}=\int d^3x\Tr\left[A\wedge dA 
+\frac{2}{3}A\wedge A \wedge A\right]$, and non-abelian helicities as the CS forms with $T^a$ inserted, e.g.,
${{\cal H}^{NA}}^a_{mm}=\Tr\left[T^a\left(A\wedge dA+\frac{2}{3}A\wedge A \wedge A\right)\right]$.

We have then constructed solutions with nonzero helicities. We have first embedded the abelian constant, plane wave, and Hopfion
solutions in the non-abelian theory. 
A radial non-abelian wave, depending on $r+t$, 
was found to have to have finite helicity, but infinite energy and to have a divergence 
at the line $\theta=0,\pi$ in spherical coordinates. A planar non-abelian wave, depending on $z+t$, also has a finite helicity, 
but infinite energy and a divergence at the plane $z+t$ fixed, $x,y$ goes to infinity. 
We have seen that we can go to a different gauge for the Dirac monopole (in the abelian case), where the helicity is finite and 
nonzero, though the energy is infinite and the field diverges at 0. But that means that in the non-abelian case, for the 't Hooft 
monopole, we could still find a different gauge for the monopole field $A_\mu^a$, where the helicity is finite, but (as usual 
for the 't Hooft monopole) the energy is finite and the fields are finite at zero and go to a constant at infinity. 
That means that this monopole solution is in some sense the analog of a Hopfion, though it is unusual. 

Finally, we have shown two ways to generalize the Bateman construction to the non-abelian case, though it is not yet clear 
which is more useful. 

There are many questions left over for future work.
\begin{itemize}
\item
 One is the construction of true non-abelian Hopfion solutions with nonzero 
helicities  {\em and Hopf index} (since it is not clear that the monopole solution that we found is such one).
\item 
 Another task is to 
consider solutions with nonzero nonabelian helicities ${{\cal H}^{NA}}^a_{r,s}$, not just singlet ones, like we have considered 
here.
\item
An interesting question is whether one can find an interpretation of the non-abelian helicities in terms of linking numbers in a similar manner to the abelian ones.\footnote{We thank C. Hoyos  for asking this question.}
\item
 We also need to understand if any of the Bateman formulation generalizations can be used to construct further, 
more complicated solutions (like for instance  $(p,q)$-knotted ones by the equivalent of the $\a\rightarrow f(\a,\b)$, 
$\b\rightarrow g(\a,\b)$ transformations, in particular $\a\rightarrow \a^p, \b\rightarrow \b^q$).
This question could be studied for the abelian theory but obviously it would me more challenging and interesting for 
non-abelian theories. 
\item 
It would also be interesting to see whether, starting with a solution with non-Abelian helicities, one can minimize the  action over 
diffeomorphisms and find an instanton-like solution, like in the Abelian case.\footnote{ We thank A. Abanov for raising this possibility.}
\item
A first attempt in formulating gravitational Hopfion following the lines of the abelian gauge theory was made in \cite{Alves:2018wku}. 
It will be interesting to push this program forward and in particular to examine possible relations with the non-abelian helicities.
\end{itemize}
We hope to come back to these questions in the future.

\section*{Acknowledgments}

We would like to thank A. Abanov and C. Hoyos for useful discussions and for  their comments on the manuscript.
The work of HN is supported in part by  CNPq grant 301491/2019-4 and FAPESP grant 2019/21281-4.
HN would also like to thank the ICTP-SAIFR for their support through FAPESP grant grant 2021/14335-0, 
and to the Physics Department at Uppsala University for its hospitality, where this article was finished.
The work of JS was supported  by a grant 01034816 titled  ``String theory reloaded- from fundamental questions to applications''     
of  the ``Planning and budgeting committee''. JS would like to thank for warm hospitality  the SCGP  at Stony Brook where this 
project has been initiated.

\bibliography{Nullnonabelian}

\bibliographystyle{utphys}

\end{document}